\documentclass[12pt]{article}
\pdfoutput=1

\usepackage[utf8]{inputenc}

\usepackage{draft}
\usepackage{putex}

 \usepackage{dsfont}
\usepackage[all]{xy}

\usepackage{stackrel,amssymb}

\usepackage{graphicx}
\usepackage{caption}
\usepackage{amsmath,bm}
\usepackage{array}
\usepackage{subcaption}
\usepackage{epstopdf}
\usepackage{enumerate}
\usepackage{cite}
\usepackage{tensor}
\usepackage{slashed}
\usepackage[utf8]{inputenc}
\usepackage{rotating}
\usepackage{bigfoot}
\usepackage[
      colorlinks=true,
      linkcolor=blue,
      urlcolor=blue,
      filecolor=black,
      citecolor=red,
      ]{hyperref}

 \usepackage{colortbl}
 
\usepackage{tikz-cd}

\def\XXint#1#2#3{{\setbox0=\hbox{$#1{#2#3}{\int}$ }
		\vcenter{\hbox{$#2#3$ }}\kern-.6\wd0}}

  \usepackage{adjustbox}
\usepackage{multirow}
\usepackage{tikz-cd}

\numberwithin{equation}{section}

\newtheorem{theorem}{Theorem}

\def\<{\langle}
\def\>{\rangle}

\def\pa{\partial}

\def\cK{\mathcal{K}}

\newtheorem{conjecture}[theorem]{Conjecture}

\newcommand{\leftrarrows}{\mathrel{\raise.75ex\hbox{\oalign{%
				$\scriptstyle\leftarrow$\cr
				\vrule width0pt height.5ex$\hfil\scriptstyle\relbar$\cr}}}}
\newcommand{\lrightarrows}{\mathrel{\raise.75ex\hbox{\oalign{%
				$\scriptstyle\relbar$\hfil\cr
				$\scriptstyle\vrule width0pt height.5ex\smash\rightarrow$\cr}}}}
\newcommand{\Rrelbar}{\mathrel{\raise.75ex\hbox{\oalign{%
				$\scriptstyle\relbar$\cr
				\vrule width0pt height.5ex$\scriptstyle\relbar$}}}}

\makeatletter
\def\leftrightarrowsfill@{\arrowfill@\leftrarrows\Rrelbar\lrightarrows}
\newcommand{\xleftrightarrows}[2][]{\ext@arrow 3399\leftrightarrowsfill@{#1}{#2}}
\makeatother

\begin{document}

\preprint{}


\title{
Casimir Energy and Modularity in  Higher-dimensional Conformal Field Theories 
}

\authors{Conghuan Luo and Yifan Wang}


\institution{NYU}{Center for Cosmology and Particle Physics, New York University, New York, NY 10003, USA}

\abstract{
An important problem in Quantum Field Theory (QFT) is to understand the structures of observables on spacetime manifolds of nontrivial topology. Such observables arise naturally when studying physical systems at finite temperature and/or finite volume and encode subtle properties of the underlying microscopic theory that are often obscure on the flat spacetime. Locality of the QFT implies that these observables can be constructed from more basic building blocks by cutting-and-gluing along a spatial slice, where a crucial ingredient is the Hilbert space on the spatial manifold. In Conformal Field Theory (CFT), thanks to the operator-state correspondence, we have a non-perturbative understanding of the Hilbert space on a spatial sphere. However it remains a challenge to consider more general spatial manifolds.
Here we study CFTs in spacetime dimensions $d>2$ on the spatial manifold $T^2\times \mR^{d-3}$ which is one of the simplest manifolds beyond the spherical topology. We focus on the ground state in this Hilbert space and analyze universal properties of the ground state energy, also commonly known as the Casimir energy, which is a nontrivial function of the complex structure moduli $\tau$ of the torus. The Casimir energy is subject to constraints from modular invariance on the torus which we spell out using $PSL(2,\mZ)$ spectral theory. Moreover we derive a simple universal formula for the Casimir energy in the thin torus limit using the effective field theory (EFT) from Kaluza-Klein reduction of the CFT, with exponentially small corrections from worldline instantons.  We illustrate our formula with explicit examples from well-known CFTs including the critical $O(N)$ model in $d=3$ and holographic CFTs in $d\geq 3$.

}
\date{}

\maketitle

\tableofcontents

\section{Introduction}

Conformal Field Theories (CFTs) provide a unifying and powerful approach to a large class of strongly coupled Quantum Field Theories (QFTs) that arise in both high energy and condensed matter physics. Over the recent years, there has been tremendous development in elucidating the structure of CFTs by the conformal bootstrap method \cite{Poland:2018epd,Poland:2022qrs,Hartman:2022zik}, which explores general constraints on the CFT observables by basic principles such as unitarity and associativity of the operator-product-expansion (OPE). In spacetime dimensions $d>2$,
most of the works focus on CFT observables defined on flat spacetime, namely the spectrum of local operators and their OPE coefficients. These CFT data determine completely the local correlation functions on the flat spacetime $\cM_d=\mR^{d-1,1}$, which are central objects in QFT. However there is much more to QFT than the flat spacetime. In particular, local QFT is expected to be defined on general spacetime manifolds.

Considerations of QFT on a spacetime manifold of nontrivial topology open the door to a profusion of new observables. For example the correlation functions of local operators (even the partition function itself)  may now depend nontrivially on topology and finer geometric data of the manifold. Such dependence 
detects global topological features of QFT which
are crucial to distinguish theories that are otherwise identical at the level of local correlation functions \cite{Aharony:2013hda,Gaiotto:2014kfa}. Of course QFT on a general spacetime manifold contains far more information than such topological data. The  challenge is to develop  methods to extract QFT observables on nontrivial manifolds, understand their general properties, and uncover  underlying structures. On the surface this seems an intractable task simply because of the sheer complexity of the geometry of the manifolds. Nonetheless, the   locality principle dictates that QFT observables must obey consistency conditions known as cutting-and-gluing axioms \cite{Atiyah:1989vu,Segal:2002ei}, which implies that the study of QFT on general manifolds can be reduced to more basic building blocks and their gluing. It also points to intricate relations among seemingly distinct observables on different manifolds.
Unsurprisingly, such questions are most well-formulated for CFTs. 
Indeed the conformal bootstrap equations for local correlation functions amount to the cutting-and-gluing consistencies on the flat spacetime. Here we decompose a CFT observable (e.g. a correlation function of local operators) into two pieces by cutting along a sphere $S^{d-1}$, and 
the basic building blocks consist of the local operator spectrum which specifies the Hilbert space of states on $S^{d-1}$ via the state-operator correspondence and the OPE coefficients that determine the wavefunction associated to each piece from the decomposition. To pursue this axiomatic approach for CFTs on $d>2$ spacetime manifolds which in general do not admit an $S^{d-1}$ spatial slice, we will need to understand the Hilbert space on non-spherical spatial manifold $\cM_{d-1}$, which is a main motivation for this work.
 
In this paper, we initiate a systematic study of $d>2$ CFTs on spatial manifold $\cM_{d-1}=T^2\times \mR^{d-3}$, which is one of the simplest non-spherical manifolds with nontrivial topology and nontrivial geometric moduli (coming from the $T^2$ factor). 
We write the metric on $T^2$ as
\ie 
ds^2_{T^2}=L_1^2 |d t + \tau d x  |^2\,,\quad \tau={L_2\over L_1} e^{i\theta}\,,
\label{T2metric}
\fe
where $L_1$ and $L_2$ are the lengths of two cycles with a tilt angle $\theta\in (0,\pi)$ and $t,x\in [0,1)$ are the periodic coordinates. In other words, the area of $T^2$ is $A=L_1 L_2 \sin \theta$ and   $\tau=\tau_1+i\tau_2$ denotes the usual shape moduli (complex structure moduli) of the torus which take values on the upper half-plane $\mH$ (i.e. $\tau_2>0$).

The immediate task is to investigate the Hilbert space $\cH_{T^2\times \mR^{d-3}}$ on  this spatial manifold, for which
little is known since there is no correspondence with local CFT operators (see comments on the $d=3$ case in \cite{Belin:2018jtf}). Here we make progress by focusing on the ground state of $\cH_{T^2\times \mR^{d-3}}$ in an ambient spacetime $\cM_d=T^2\times \mR^{d-3,1}$
and exploring general properties of the ground state energy, also known as the Casimir energy,\footnote{See \cite{Dantchev:2022hvy} for a recent review on critical Casimir effect in general.}  measured with respect to the Hamiltonian 

\ie 
H_{T^2\times \mR^{d-3}}=\int_{T^2 \times \mR^{d-3}} T_{00}\,.
\fe
The Casimir energy is extensive in the noncompact $\mR^{d-3}$ directions, whose energy density we denote by $E_{\rm vac}$ takes the following form
\ie 
E_{\rm vac}=-{\cE(\tau)\over A^{d-2\over 2}}\,,
\label{Evactodensity}
\fe
where the dependence on the torus area $A$ is fixed by dimensional analysis and the dimensionless coefficient $\cE(\tau)$ will be the central object in this paper that contains nontrivial dependence of the shape moduli $\tau$ of the torus. We will refer to $\cE(\tau)$ as the (dimensionless) Casimir energy density.  We emphasize that $E_{\rm vac}$ has a counterterm ambiguity that is fully-extensive on $T^2\times \mR^{d-3}$ and shifts $E_{\rm vac} \to E_{\rm vac} + A\Lambda^d$ where $\Lambda$ is a UV cutoff scale. This comes from the cosmological constant counterterm on $T^2\times \mR^{d-3,1}$. There are no other counterterm ambiguities since $T^2$ is flat. Consequently, $\cE(\tau)$ is unambiguous and well-defined.

We note the following immediate properties of $\cE(\tau)$. Firstly, $\cE(\tau)$ is a real function of the complex moduli $\tau$ since $H_{T^2\times \mR^{d-3}}$ is Hermitian. Secondly, 
 for bosonic CFTs, $\cE(\tau)$ is modular invariant under $PSL(2,\mZ)$ transformations of $\tau$,
\ie 
\cE(\tau)=\cE (\C \tau)\,,\quad \C\tau ={a\tau+b\over c\tau+d}\,,
\fe
with $a,b,c,d\in\mZ$ and $ad-bc=1$. This follows from the orientation-preserving large diffeomorphisms on $T^2$ and is also a general property of the Hamiltonian $H_{T^2\times \mR^{d-2}}$ and its entire energy spectrum.\footnote{For fermionic CFTs, to define the theory on $T^2$, a choice of spin structure must be specified. This breaks the $PSL(2,\mZ)$ modular invariance to congruence subgroups. We study the toroidal Casimir energy density of fermionic CFTs in a subsequent publication \cite{FermionCFT}.} Thirdly, if the CFT is parity-invariant then 
\ie 
\cE(\tau)=\cE(-\bar\tau)\,,
\fe
which is a consequence of the orientation-reversing large diffeomorphism $(x_1,x_2)\to (x_1,-x_2)$ on $T^2$.

To derive further universal constraints on $\cE(\tau)$, we consider the Euclidean CFT on $T^2\times \mR^{d-2}$ (see previous works \cite{Shaghoulian:2015kta,Belin:2016yll}). We first take one of $\mR^{d-2}$ directions to be the Euclidean time, then the partition function is clearly determined by the Casimir energy. Alternatively, choosing the cycle $S^1_\B$ on $T^2$ of length $\B=L_1$ to be the Euclidean time, the partition function becomes a trace over the Hilbert space on the base $S^1\times \mR^{d-2}$. This second perspective enables us to infer the structure of $\cE(\tau)$ in the limit $\tau\to i\infty $ where $L_1$ is small compared to $L_2$ using the effective field theory (EFT) of the CFT from Kaluza-Klein (KK) reduction on $S^1_\B$. A main result is the following universal behavior for the Casimir energy density,\footnote{Despite the funny looking powers in $\tau_2$ (which come from the rescaling \eqref{Evactodensity}), the physical interpretations of the perturbative terms are quite simple: $c_1$ controls the extensive dependence of the CFT Casimir energy on the large circle of size $L_2$ while $c_2$ controls the sub-extensive dependence on $L_2$.} 
\ie 
\lim_{\tau \to i\infty } \cE(\tau)=
c_1 \tau_2^{d\over 2}
+c_2 \tau_2^{1-{d\over 2}}
+\cO \left(\tau_2^\A e^{-2\pi n M\tau_2} e^{2\pi i n Q \tau_1 }\right) \,,
\label{cuspform}
\fe
where $n>0$ and $Q$ are integers and $c_1,c_2,\A,M$ are theory-dependent constants.
Namely the cusp behavior of $\cE(\tau)$ is controlled by two perturbative terms in $\tau_2$ 
and a tower of non-perturbative (exponentially small) terms which may also come with perturbative corrections. As we explain in Section~\ref{sec:EFT}, the constants $c_1,c_2$ are captured by the EFT from integrating out massive KK modes and the absence of other perturbative terms follows from a simple EFT argument. On the other hand, 
the non-perturbative terms in \eqref{cuspform} come from worldline instantons from massive KK particles with (rescaled) mass $M$ and KK charge $Q$ going $n$-times along the $S^1$ on the base of the KK reduction.

Moreover we will argue that the leading perturbative coefficient $c_1\geq 0$ and is strictly positive unless the theory is topological, thus providing a measure of local degrees of freedom in the $d$-dimensional CFT. Similarly, the second perturbative coefficient $c_2\geq 0$ is a measure of the gapless degrees of freedom in the $d-1$-dimensional  EFT from KK reduction, and $c_2=0$ if the EFT is gapped. 
 


The perturbative coefficients $c_1,c_2$ in \eqref{cuspform} are closely related to CFT observables at finite temperature (in $d$ and $d-1$ dimensions respectively). Such observables are of great interest in both condensed matter and high energy literature because they encode finite temperature properties of quantum critical points \cite{Sachdev:1993pr,Chubukov_1994,Katz:2014rla} and via the AdS/CFT correspondence provide a CFT description of black holes \cite{Maldacena:1997re,
Gubser:1998bc,Witten:1998qj,Witten:1998zw}.
The relevant spacetime manifold here is $\cM_d=S^1_\B\times \mR^{d-1}$ for the $d$-dimensional CFT, and the temperature is related to the inverse size of the circle $T={1\over \B}$.\footnote{The thermal background  $\cM_d=S^1_\B\times \mR^{d-1}$ is one of the simplest manifolds that is not conformally flat. Nonetheless all known observables on $\cM_d=S^1_\B\times \mR^{d-1}$ are still determined by the flat space CFT data \cite{Iliesiu:2018fao}.}
The basic observables here are the one-point functions of conformal primaries which determine higher-point functions by OPE. In contrast to the flat spacetime, the one-point function on the thermal background is in general non-vanishing and determined by the residual conformal symmetry up to an overall constant \cite{Iliesiu:2018fao}. 
A distinguished case is the one-point function of the stress energy tensor
\ie 
\la T_{tt} \ra_{S^1_\beta \times \mR^{d-1}}=(1-d)\la T_{ii} \ra_{S^1_\beta \times \mR^{d-1}}= {(d-1)b_T\over d\B^d}\,,\quad  \la T_{ti}  \ra_{S^1_\beta \times \mR^{d-1}}= 0\,,
\label{Tmn1PF}
\fe
where $t$ labels the $S^1_\B$ direction and $i$ labels the spatial $\mR^{d-1}$ directions. 
The same coefficient $b_T$ determines the free energy density 
\ie 
f(\B)= - {1\over \B V_{d-1}}\log Z_{S^1_\B\times \mR^{d-1}}={b_T\over  d \B^{d}}\,,
\label{fbeta}
\fe 
where $V_{d-1}$ regulates the infinite spatial volume. This follows from \eqref{Tmn1PF} and the relation
\ie 
\la T_{tt} \ra_{S^1_\beta \times \mR^{d-1}} = {1\over V_{d-1}}\pa_\B \log Z_{S^1_\B\times \mR^{d-1}}\,. 
\fe
The thermal background can be thought of a limit of $\cM_d=S^1_\B\times S^{d-1}_R$
with ${R\over \B} \to \infty$. The latter comes naturally from the radial quantization of the CFT on the sphere $S^{d-1}_R$ of radius $R$. The corresponding partition function counts states in the Hilbert space $\cH_{S^{d-1}}$ graded by their energy measured by the radial Hamiltonian (equivalently local operators graded by their scaling dimensions $\Delta$),
\ie 
Z_{S^1_\B\times S^{d-1}_R}\equiv\Tr_{\cH_{S^{d-1}}} e^{-{\B\over R} \Delta }\quad \xrightarrow[]{{R\over \B} \to \infty} 
\quad 
Z_{S^1_\B\times \mR^{d-1}} =  \exp \left(-{ {\rm vol}(S_R^{d-1})  \over d \B^{d-1}}b_T \right)\,,
\fe
which is related to the thermal partition function on $S^1_\B \times \mR^{d-1}$ in the ${R\over \B} \to \infty$ limit where the heavy operator contributions are no longer suppressed. 
Therefore 
$b_T$ measures the asymptotic  density of high dimension local operators, akin to the role of the conformal central charge $c_{2d}$ in $d=2$ CFT,\footnote{We emphasize that in $d>2$, the coefficient $b_T$ is in general a nontrivial function on the conformal manifold of the CFT.} and we see that the finite temperature one-point function $\la T_{\m\n}\ra_{S^1_\B \times \mR^{d-1}}$ is determined by the flat space operator data.\footnote{Similarly, the one-point function of a general local operator $\la \cO \ra_{S^1_\B \times \mR^{d-1}}$ is related to the ${R\over \B} \to \infty$ limit of $\la \cO(x) \ra_{S^1_\B \times S^{d-1}_R}$,
\ie 
\la \cO(x) \ra_{S^1_\B \times S^{d-1}_R} = \sum_{\phi} e^{-{\B\over R} \Delta_{\phi}} \la \phi |\cO(x) |\phi\ra \,,
\fe
which depends on the OPE coefficients with general CFT operators $\phi$.}


Coming back to the coefficients $c_1$ and $c_2$ in our universal formula for the Casimir energy \eqref{cuspform}, as we explain in Section~\ref{sec:EFT}, $c_1$ is determined by the thermal one-point function of the CFT stress energy tensor,\footnote{In the notation of \cite{Iliesiu:2018fao}, $c_1^{\rm here}=-f^{\rm there}$.}
\ie 
c_1= -{b_T\over d}\,.
\fe
This is natural because the $\cM_d=T^2\times \mR^{d-2}$ background we consider is
a generalization of the thermal background by including another compact direction which becomes large in the limit of \eqref{cuspform}.

If the $d$-dimensional CFT upon reduction on $S^1$ contains a gapless sector described by a $(d-1)$-dimensional CFT with stress energy tensor $\widehat T$, $c_2$ is determined by the corresponding  thermal one-point function in $d-1$ dimensions,
 \ie 
c_2= -{b_{\widehat T}\over d-1}\,,
\fe
which no longer has an obvious interpretation in terms of flat space operator data in $d$-dimensions.

The rest of the paper is organized as follows. In Section~\ref{sec:genstructureEFT}, we carry out the effective field theory analysis to derive the universal behavior of the Casimir energy density $\cE(\tau)$ in the thin torus limit. In Section~\ref{sec:spectraltheory}, 
we spell out the modular properties of  $\cE(\tau)$ using $PSL(2,\mZ)$ spectral theory.
We then illustrate our formula \eqref{cuspform} with concrete CFT examples, starting from the free scalar theory in $d>2$ in Section~\ref{sec:free} where the Casimir energy $\cE(\tau)$ is given by familiar real analytic Eisenstein series. For interacting CFTs, $\cE(\tau)$ is much harder to compute exactly. In Section~\ref{sec:ON}, we study the $d=3$ critical $O(N)$ CFT in the large $N$ limit on $T^2\times \mathbb{R}$ extending previous works. We derive new formulae for $\cE(\tau)$ in the $O(N)$ CFT in terms of a generalized Eisenstein series and determine the behavior in the thin torus limit which matches onto our formula \eqref{cuspform} in an interesting way. In large $N$ CFTs with a semi-classical holographic dual, the ground state on $T^2\times \mR^{d-3,1}$ is dual to certain AdS soliton in the bulk which depends on the moduli $\tau$ of the boundary $T^2$. The Casimir energy density $\cE(\tau)$ is then determined by the bulk action evaluated on the AdS soliton solution which we discuss in Section~\ref{sec:holo}.
We end with a discussion of open questions and future directions in Section~\ref{sec:discussion}.


\section{General Structure of Casimir Energy Density} 
\label{sec:genstructureEFT}

\subsection{High Temperature Expansion from Effective Field Theory}
\label{sec:EFT}

Here we derive the universal formula \eqref{cuspform} for the Casimir energy density $\cE(\tau)$ in the thin torus limit $\tau\to i\infty$. As already mentioned in the introduction, this will be achieved by analyzing the effective field theory (EFT) from KK reduction on the  small circle in this limit \cite{Banerjee:2012iz,DiPietro:2014bca}.

We work with the CFT on $T^2\times \mR^{d-2}$ and start by putting the metric \eqref{T2metric} into the KK form
\ie 
ds^2=\beta^2(dt+a)^2+ L^2 dx^2+ dy^2_{\mathbb{R}^{d-2}}\,,
\label{KKmetric}
\fe 
where $a=\mu dx$ is KK photon (graviphoton) and 
\ie 
\B=L_1\,,\quad \m=\tau_1\,,\quad L=L_2\sin\theta=\tau_2\B\,.
\label{changevar}
\fe
Treating the $t$ direction as Euclidean time, we can interpret the CFT partition function on $T^2\times \mR^{d-2}$ as a thermal partition function over the Hilbert space on the base manifold $S^1_L\times \mR^{d-2}$,
\ie 
Z(T^2\times \mR^{d-2})=\Tr_{\cH_{S^1_L\times \mR^{d-2}}} \left[   e^{-\B (H_{S^1_L\times \mR^{d-2}}+i \m P_{S^1_L\times \mR^{d-2}}) } 
\right]\,,
\fe
where $P_{S^1_L\times \mR^{d-2}}$ is the generator for translation along the base $S^1_L$ (the charges are quantized as ${n\over L}$ for $n\in \mZ$) and $\m$ the corresponding chemical potential.  

Since we can equivalently pick one of the non-compact directions in $\mR^{d-2}$ to be Euclidean time, in which case the partition function is dominated by the ground state contribution,
we arrive at the following relation 
\ie 
e^{-V_{d-2}E_{\rm vac}(L_1,L_2,\theta)}= e^{-\B F_{S^1_L\times \mR^{d-2}}(L,\B,\m) }
\fe
between the ground state energy on $T^2\times \mR^{d-3}$ and the free energy on $S^1_L\times \mR^{d-2}$. From the extensivity of the free energy and \eqref{Evactodensity} with $A=L\B$ (the area of $T^2$),  we obtain a relation 
\ie 
\cE(\tau) = -A^{d\over 2} f(L, \B,\m)\,,
\label{EtoF}
\fe
between the Casimir energy density $\cE(\tau)$ on $T^2\times \mR^{d-3}$ and the free energy density $f(\B,\m)$ on $S^1_L\times \mR^{d-2}$. 

We now focus on the limit $\B \ll L$ with $\m$ fixed (equivalently $\tau_2\to \infty$ with $\tau_1$ fixed). In this case, $f(\B,\m)$ is naturally captured by the EFT from KK reduction of the CFT on $S^1_\B$ in \eqref{KKmetric}. Below we will deduce universal properties of $f(\B,\m)$ from general EFT considerations.

On general grounds, the $d-1$ dimensional Wilsonian EFT from the KK reduction of a $d$ dimensional CFT on $S^1_
\B$ is obtained from integrating out KK modes of mass $m_{\rm KK}\sim {1\over \B}$. At infinite volume (on the base manifold), it is well-known that this leads to an effective action $S_{\rm EFT}$ that consists of local analytic functionals in the background fields which capture the induced contact interactions. Here the relevant background fields consist of the metric on the base manifold $g_{ij}$ and the KK photon gauge field $a_i$. The possible local functionals are constrained by invariance under background diffeomorphism and gauge transformations (together they correspond to diffeomorphism transformations in $d$ dimensions). These functionals take the form  
\ie
S_{\rm EFT} \ni \B^{2n+2m-d+1} \int d^{d-1}x\sqrt{g}R[g]^n F[a]^m\,,
\label{localfunctional}
\fe
for $n,m\in \mZ_{\geq 0}$ where $R[g]$ denotes schematically the Riemann curvature on the base manifold and $F[a]$ the field strength of the KK photon. They constitute a derivative expansion of $S_{\rm EFT}$ where the $\B$ dependence is fixed by dimensional analysis, and so equivalently describes a high temperature (small $\B$) expansion of $S_{\rm EFT}$.

Now the background we are interested in \eqref{KKmetric} has vanishing curvature for both the metric $g_{ij}$ and the gauge field $a_i$ on the base. Consequently, the only non-vanishing local functional among \eqref{localfunctional} has $n=m=0$, which corresponds to the cosmological constant term generated by integrating out massive KK modes.
Therefore, we have 
\ie 
S_{\rm EFT}=- {c_1\over \B^{d-1}} \int d^{d-1}\sqrt{g}   + S_{\rm gapless} \,,
\label{SEFT}
\fe
where $S_{\rm gapless}$ is 
the action that governs the dynamics of the gapless modes that survive the KK reduction. Note that quantum effects at finite temperature typically generate a non-negative thermal mass squared $m^2_{\rm thermal}$ for the KK zero mode. While there have been recent counter-examples to this statement (see \cite{Chai:2020onq,Chaudhuri:2021dsq}), here we assume this is true. The remaining unlifted KK zero modes are then described by a $d-1$ dimensional CFT.

An important caveat here is that if the $d$ dimensional CFT has gravitational anomalies, such anomalies must be matched by the EFT. In this case, local functionals of the Chern-Simons type  can appear in $S_{\rm EFT}$, which are generally not invariant under background gauge transformations\footnote{Such Chern-Simons type terms are either non-invariant under small gauge transformations because of field-dependent couplings
or non-invariant under large gauge transformations because of ill-quantized constant couplings.} and the non-invariance precisely reproduces the $d$ dimensional anomalous variation upon reduction on $S^1_\B$. Nevertheless, the flatness of our background ensures that they do not contribute to $S_{\rm EFT}$.\footnote{This is to be contrasted with the cases where the base manifold of the KK reduction has nontrivial curvature (e.g. $S^{d-1}$ or $S^{1}\times S^{d-2}$ for $d\geq 4$).}

We emphasize that the simple form of the $S_{\rm EFT}$ in \eqref{SEFT} completely determines the EFT partition function, correspondingly the free energy density $f(\B,\mu)$, in the high temperature limit to all orders in $\B$\,,
\ie 
\lim_{\beta \to 0} \,f(\beta,\mu)={ -{c_1\over \beta^{d}}} + {1\over \beta}\left(  -{{c_2\over L^{d-1}} } \right)+ f_{\rm np}(\beta,\mu)\,,
\label{highTf}
\fe 
up to terms that are non-perturbative in $\beta$ (i.e. exponentially small as $\beta \to 0$) which we package together in $f_{\rm np}(\beta,\mu)$ and will come back to shortly. The first term on the RHS of \eqref{highTf} comes obviously from the cosmological constant term in \eqref{SEFT}. Comparing with \eqref{fbeta} in the $\beta\ll L$ limit, we find
\ie 
c_1=-{b_T\over d}
\label{c1rel}
\fe
is determined by one-point function of the stress tensor $T_{\m\n}$.
The second term  on the RHS of \eqref{highTf} comes from the thermal free energy of the $d-1$ dimensional CFT (that survives the KK reduction) on $S^1_L\times \mR^{d-2}$ (i.e. \eqref{fbeta}  with $\beta$ replaced by $L$ and $d$ replaced by $d-1$) and the coefficient $c_2$ is determined by the one-point function of the stress tensor $\widehat T_{ij}$ for  this $d-1$ dimensional CFT,
\ie 
c_2=-{b_{\widehat T}\over d-1}\,.
\label{c2rel}
\fe

\subsection{Non-perturbative Corrections from Worldline Instantons}
\label{sec:worldlineinstanton}

The high temperature EFT does not completely determine the free energy of the KK-reduced theory on the base manifold away from the infinite volume limit. 
The corrections are denoted by $f_{\rm np}(\B,\m)$ in \eqref{highTf} which are exponentially suppressed (i.e. behave as $e^{-L/\B}$). Such non-perturbative contributions are naturally associated with worldline instantons, namely virtual massive particles propagating around the compact circle $S^1_L$ on the   base manifold. They are most easily seen from the worldline representation for the one-loop effective action of a free massive particle \cite{Strassler:1992zr,Dunne:2005sx},\footnote{Here for simplicity we focus on scalar particles. Generalizations to spinning particles can be found in \cite{Strassler:1992zr} by introducing auxiliary variables along the worldline.} 
 \ie 
S_{\tilde M,Q}=\int_0^\infty {dT\over T} e^{-{  \tilde M^2} T} \int_{x^j(T)=x^j(0)} D x^j(\varphi) 
\exp\left[-\int_0^T d\varphi \left({1\over 4} g_{ij}\dot x^i \dot x^j+  2\pi i   Q  a_j  \dot x^j \right)\right]\,,
\label{SMQ}
\fe 
where $x^j(\varphi)$ parametrizes  the worldline of a scalar particle of mass $\tilde M\sim {1\over \B}$ and KK-charge $Q\in \mZ$ and $T$ is the usual Schwinger parameter. For nonzero $Q$, the particle couples to the background KK-photon $a_i$. 
The saddle point equation for the worldline path integral over $x^j(\varphi)$ is simply $\ddot x^i=0$ since the $a_i$ is flat. The Euclidean solutions compatible with the periodicity condition are
\ie 
x^i(\varphi)=({n \varphi\over T},0,\dots,0)\,,
\label{wsinst}
\fe
where the only nonzero entry is in the $S^1_L$ direction and this solution winds $n$ times along $S^1_L$.  The one-loop effective action \eqref{SMQ} is dominated by these saddles for large mass $\tilde M L \sim {L/\B} \gg 1$,
\ie 
\lim_{\tilde M L \gg 1} S_{\tilde M,Q} = e^{-n \tilde M L}e^{2\pi i n Q\m }\,,
\label{wsinstact}
\fe
on top of which there are perturbative contributions in $1/\tilde M$ coming from the fluctuations.  

The $n=0$ case of \eqref{wsinst} is the trivial saddle which is insensitive to compact directions of the spacetime manifold. The fluctuations thereof precisely generate the perturbative terms in the effective action that are power law in $\beta$. The   case $n\neq 0$ of \eqref{wsinst} corresponds to nontrivial worldline instantons and \eqref{wsinstact} gives the instanton action which depends on the chemical potential $\m$ if the particle carries a nonzero KK charge $Q$. For  convenience we define the dimensionless mass $M$ by $\tilde M={2\pi M\over \B}$.

The most general contributions to $f_{\rm np}(\B,\m)$ come from multiple worldline instantons, each labeled by its mass $M$, KK charge $Q$ and winding number $n$. Moreover these worldline instantons may interact with one another, producing corrections beyond the dilute instanton gas approximation \cite{coleman1988aspects}.

The non-perturbative contributions to the free energy takes the following form, as a sum over multi-instanton configurations
\ie
A^{\frac{d}{2}} f_{\rm np}(\beta,\mu) = -\sum_{j=1}^\infty   \sum_{\vec M,\vec Q}
e^{-{2\pi  L\over \B} \sum_{i=1}^j M_i }e^{2\pi i \m  \sum_{i=1}^j Q_i } g_{j,\vec M,\vec Q}\left({\B\over L}\right)\,.
\label{fnpstruct}
\fe
Here $j$ counts the number of worldline instantons each with unit winding number. The vectors $\vec M,\vec Q$ each has $j$ entries and they specify the masses and KK charges of individual  instantons in this configuration.\footnote{In this parametrization, a worldline instanton of charge $Q$ and winding number $n$ is represented as $n$ identical worldline instantons of unit winding number and the same charge $Q$.}
The fluctuations around each $j$-instanton configuration and the integration over the moduli space of such configurations are captured by the factor $ g_{j,\vec M,\vec Q}\left({\B\over L}\right)$ which contains power-law terms in ${\B\over L}$.

For free CFTs compactified on $S^1_\B$, the worldline instantons 
coming from massive KK modes do not interact and  $ g_{j,\vec M,\vec Q}$ can be computed explicitly from \eqref{SMQ}. See further discussions in Section~\ref{sec:free} for the free scalar CFT. 
 In general CFT upon circle reduction, there are nontrivial interactions among worldline instantons which are not taken into account by \eqref{SMQ}. In Section~\ref{sec:ON}, we will find explicit expressions of $g_{j,\vec M,\vec Q}$ for the worldline instantons in the interacting $O(N)$ CFT.

Using the relation between the free energy and the Casimir energy \eqref{EtoF} (and also \eqref{changevar}), we conclude the EFT analysis with the following universal behavior for the Casimir energy density on $T^2\times \mR^{d-3,1}$,
\ie 
\lim_{\tau \to i\infty } \cE(\tau)=
c_1 \tau_2^{d\over 2}
+
c_2 \tau_2^{1-{d\over 2}}
+\sum_{j=1}^\infty   \sum_{\vec M,\vec Q}
e^{-{2\pi  \tau_2} \sum_{i=1}^j M_i }e^{2\pi i \tau_1 \sum_{i=1}^j Q_i } g_{j,\vec M,\vec Q}(\tau_2)\,,
\label{EFTres}
\fe
which contains two perturbative terms whose coefficients are related to finite temperature CFT data by \eqref{c1rel} and \eqref{c2rel}, and a tower of nonperturbative contributions from the worldline instanton gas.

\section{Modular Properties from $PSL(2,\mZ)$ Spectral Theory}
\label{sec:spectraltheory}

As already explained in the introduction, the toroidal Casimir energy density $\cE(\tau)$ is a modular (invariant) function of the complex moduli $\tau$. 
Here we want to understand what types of non-holomorphic modular functions are relevant for this CFT observable.

Unlike the holomorphic case where the space of modular functions is highly constrained (to be rational functions of the modular invariant $j(\tau)$), the space of non-holomorphic modular functions is much bigger. This is necessary to accommodate the vast zoo of modular functions of different properties that arise from QFT observables  which are mostly non-holomorphic. Well-studied examples of such modular functions include the torus partition functions of $d=2$ CFTs and correlation functions in the $d=4$ $\cN=4$ super-Yang-Mills (SYM) CFT where $\tau$ is the complexified Yang-Mills coupling. An important question there is to understand the general properties of the relevant modular functions as dictated by the underlying CFT. Impressive progress is made in the recent years on this question for both the $d=2$ CFTs \cite{Benjamin:2021ygh,Benjamin:2022pnx} and the $d=4$ SYM \cite{Collier:2022emf} using the powerful tool of $PSL(2,\mZ)$ spectral theory (see also previous work \cite{Green:2014yxa}). Here we initiate to tackle the same question in yet another different context, that is the toroidal Casimir energy of general $d>2$ CFTs.

\subsection{Review of $PSL(2,\mZ)$ Spectral Theory} 
Let us start with a quick review of the necessary ingredients for the $PSL(2,\mZ)$ spectral theory. See for example \cite{iwaniec2002spectral,terras2013harmonic} for more details. Here we follow the conventions of mathematics literature to denote the real and imaginary parts of the complex moduli as $\tau=x+iy$ (which we haven written as $\tau=\tau_1+i\tau_2$ in the rest of the paper).
The standard $PSL(2,\mZ)$ fundamental domain is denoted by $\cF=PSL(2,\mZ)\backslash\mH$ with the hyperbolic metric
\ie 
ds^2={dx^2+dy^2\over y^2}\,,
\fe
and hyperbolic volume
\ie 
{\rm vol}(\cF)={\pi\over 3}\,.
\fe
For modular invariant square integrable functions $f(\tau),g(\tau) \in L^2(\cF)$, there is a natural inner product known as the Petersson inner product defined by the following integral over $\cF$,
\ie 
\la f,g \ra \equiv \int_\cF {d^2 \tau \over y^2} f(\tau) \overline{g(\tau)}\,,
\label{Piprod}
\fe
Square integrability here simply means finite $\la f,f\ra$.
The hyperbolic Laplacian on $\cF$  defined by 
\ie 
\Delta \equiv -y^2(\pa_x^2+\pa_y^2)\,,
\fe
is self-adjoint with respect to \eqref{Piprod} on $L^2(\cF)$. Consequently, any modular invariant function $f\in L^2(\cF)$ admits a unique orthogonal decomposition into eigenfunctions with respect to $\Delta$,
\ie 
f(\tau)=\sum_{j=0}^\infty \n_j(\tau) {\la f,\n_j\ra \over \la \n_j,\n_j\ra}
+{1\over 4\pi i}\int_{{\rm Re\,} s={1\over 2}} ds 
\la f,E_s\ra  E_s(\tau)\,, 
\label{RSdecomp}
\fe
known as the Roelcke-Selberg spectral decomposition. The eigenfunctions here consists of a continuous family of non-holomorphic Eisenstein series $E_s(\tau)$ with $s={1\over 2}+i\mR$, which satisfies
\ie 
\Delta E_s =s(1-s) E_s\,,
\label{EisenLaplace}
\fe
and a
discrete family called the cusp Maass forms $\n_i(\tau)$
satisfying 
\ie 
 \Delta  \n_j= \lambda_j \n_j\,,\quad \lambda_j={1\over 4} + R_j^2\,,
\fe 
where $R_j$ are a set of sporadic positive real numbers. 

Let us review some properties of these modular eigenfunctions below. 
The Eisenstein series has the following explicit definition, in the form of a Fourier decomposition,\footnote{Note that the Eisenstein series is not in $L^2(\cF)$. Nonetheless when smeared by a function $f(t)\in L^2(\mR)$, $\int_\mR dt\, f(t) E_{1/2+it}(\tau)$ defines modular functions in $L^2(\cF)$.}
\ie 
E_s(\tau)=\varphi_s(y)+\sum_{j=1}^\infty 4 \cos(2\pi j x) {\sigma_{2s-1}(j)\over j^{s-1/2} \Lambda(s)} \sqrt{y} K_{s-{1\over 2}}(2\pi j y)\,,\quad 
\varphi_s(y)=y^s +{\Lambda(1-s)\over \Lambda(s)} y^{1-s}\,,
\label{genEis}
\fe
where $\sigma_m(n)=\sum_{d|n} d^m$ is the divisor function and
$\Lambda(s)$ is the completed Riemann zeta function defined below,
\ie 
\Lambda(s) \equiv \pi^{-s} \Gamma(s) \zeta(2s)\,.
\label{Lambdadef}
\fe
It satisfies the following reflection relation,
\ie 
\Lambda(s)=\Lambda\left({1\over 2}-s \right)\,,
\fe
and is meromorphic for $s\in \mC$ with simple poles at $s=0,{1\over 2}$ and residues
\ie 
{\rm Res}_{s=0} \Lambda(s) =-{\rm Res}_{s={1\over 2}} \Lambda(s) =-{1\over 2}\,,
\label{LFres}
\fe 
$\Lambda(s)$ has zeroes at $s={\rho \over 2}$ for nontrivial zeros $\rho$ of the Riemann zeta function (with ${\rm Re}\, \rho= {1\over 2}$ according to the Riemann hypothesis). 


Correspondingly, the Eisenstein series $E_s(\tau)$ is meromorphic in $s$, with a simple pole at $s=1$ and other sporadic simple poles at $s={\rho \over 2}$. In particular the residue at $s=1$ is,
\ie 
{\rm Res}_{s=1} E_s(\tau)={3\over \pi} ={\rm vol}(\cF)^{-1}\,.
\fe
For ${\rm Re}\, s > 1$, the Eisenstein series is equivalently  defined by the following Poincar\'e series
\ie 
E_s(\tau) =\sum_{\C \in \Gamma_\infty\backslash PSL(2,\mZ) } {\rm Im} (\C \tau)^s =   {1\over 2}\sum_{m,n\in \mZ,(m,n)=1}  {y^s\over |m\tau+n|^{2s}}\,,
\label{EsPsdef}
\fe
which is clearly $PSL(2,\mZ)$ invariant and its meromorphic continuation to $s\in \mC$ is given by \eqref{genEis}.  
It is convenient to define the completed (symmetric) Eisenstein series
\ie 
E_s^*(\tau)\equiv \Lambda(s) E_s(\tau)\,,
\fe
which satisfies
\ie 
E_s^*(\tau)=E^*_{1-s} (\tau)\,.
\label{Esreflection}
\fe
For ${\rm Re}\, s>1$, it can also be written as
\ie 
E^*_s(\tau) ={1\over 2} \pi^{-s} \Gamma(s)  \sideset{}{'}\sum_{m,n\in \mZ}  {y^s\over |m\tau+n|^{2s}}\,,
\label{Essym}
\fe
where $\sum'$ means the $m=n=0$ term is dropped. 
The overlaps in \eqref{RSdecomp} are explicitly given by,
\ie 
\la F ,E_s\ra
=\int_\cF {d^2\tau \over y^2} F(\tau) E_{1-s}(\tau)=\int_0^\infty dy y^{-1-s} F_0(y)\,,
\fe 
where $F_0\equiv \int_{-1/2}^{1/2} dx F(\tau)$ is the zero Fourier mode of $F(\tau)$ and the last equality follows from the unfolding trick (using the Poincar\'e series definition of $E_s$ in \eqref{EsPsdef} and then analytic continuation).

The Maass cusp forms $\n_j(\tau)$ are eigenfunctions of $\Delta$ in $L^2(\cF)$. The special case is the constant function usually normalized as,
\ie 
\n_0\equiv \sqrt{3\over \pi}\,. 
\fe
For more general Maass cusp forms, depending on their parity under $x\to -x$, we have
\ie 
\n_j ^+(\tau)=&
\sum_{k=1}^\infty 
a_k^{(j,+)}\cos (2\pi k x) \sqrt{y} K_{i R_j^+} (2\pi k y)\,,
\\
\n_j^-(\tau)=&
\sum_{k=1}^\infty 
a_k^{(j,-)}\sin (2\pi k x)\sqrt{y} K_{i R_j^-} (2\pi k y)\,,
\label{cuspforms}
\fe
where the order of the Bessel functions are related to the $\Delta$ eigenvalue $\lambda_j$ by 
\ie 
\lambda_j^\pm\equiv {1\over 4}+(R_j^\pm )^2\,.
\fe
The normalization is commonly fixed by $a_1^{(j,\pm)}=1$. The coefficients $a_j^{(j,\pm)}$ and $R_j^\pm$ are not known analytically but numerical results are available (see for example Appendix A.2 of \cite{Benjamin:2021ygh}). 

\subsection{Spectral Decomposition of Casimir Energy Density}

The spectral decomposition \eqref{RSdecomp} only applies to modular functions that are square integrable. As we have derived from EFT arguments in Section~\ref{sec:EFT},
the CFT toroidal Casimir energy density $\cE(\tau)$ has the following cusp behavior (see \eqref{EFTres}), 
\ie 
\lim_{\tau \to i\infty} \cE(\tau)=c_1 y^{d\over 2}
+c_2 y^{1-{d\over 2}} +\cO\left(y^\A e^{-2\pi M n y} e^{2\pi i kn x}\right)\,,
\label{Ecuspbehavior}
\fe
for $c_1>0$ and $c_2\geq 0$, thus clearly not square integrable for $d>2$.

Nevertheless there is a natural way to regularize this cusp behavior using the hyperbolic Laplacian,
\ie 
\cE_{\rm reg}(\tau)\equiv \left(\Delta +{d(d-2)\over 4}\right) \cE\,,
\label{diffreg}
\fe
which is modular invariant and vanishes exponentially at the cusp
\ie 
\lim_{\tau\to i\infty} \cE_{\rm reg}(\tau) =\cO\left(y^\A e^{-2\pi M n y} e^{2\pi i kn x}\right)
\fe
and thus now in $L^2(\cF)$. Therefore we can apply the decomposition \eqref{RSdecomp} and write,  
\ie 
\cE_{\rm reg}(\tau)
=\sum_{j=0}^\infty{\la \cE_{\rm reg},\n_j\ra \over \la \n_j ,\n_j\ra } \n_j(\tau)
+{1\over 4\pi i } \int_{{\rm Re\,}s={1\over 2}} ds {\la \cE_{\rm reg},E_s\ra } E_s(\tau) \,.
\fe
It is straightforward to invert the differential operator in \eqref{diffreg} and obtain,
\ie 
 \left ( \Delta +{d(d-2)\over 4} \right)^{-1} \cE_{\rm reg}
=\sum_{j=0}^\infty{\la \cE_{\rm reg},\n_j\ra \over \lambda_j +{d(d-2)\over 4}} {\n_j(\tau) \over \la\n_j,\n_j\ra}
+{1\over 4\pi i } \int_{{\rm Re\,}s={1\over 2}} ds {\la \cE_{\rm reg},E_s\ra \over {(d-1)^2\over4}-{(2s-1)^2\over 4}} E_s(\tau) \,,
\fe
which is still in $L^2(\cF)$ since coefficients in the decomposition are now divided by positive numbers that are bounded below by ${(d-1)^2\over 4}$.

This inversion is ambiguous up to the kernel of $\Delta+{d(d-2)\over 4}$ which can be fixed by adding the Eisenstein series $E_{d\over 2}(\tau)$ whose coefficient is fixed by the cusp behavior of $\cE$ in \eqref{Ecuspbehavior}. 
We thus find\footnote{In parity-invariant CFT, only the even cusp forms $\n_j^+$ in \eqref{cuspforms} appears in the spectral decomposition of the toroidal Casimir energy.}
\ie 
\cE=&\,{c_1 }E_{d\over 2}(\tau)+
\sum_{j=0}^\infty{\la \cE_{\rm reg},\n_j\ra \over \lambda_j +{d(d-2)\over 4}} {\n_j(\tau)\over \la \n_j,\n_j\ra}
+{1\over 4\pi i } \int_{{\rm Re\,}s={1\over 2}} ds {\la \cE_{\rm reg},E_s\ra \over {(d-1)^2\over4}-{(2s-1)^2\over 4}} E_s(\tau)\,.
\fe 
This decomposition is unique. Suppose there is another modular function $\cE'$ with the same cusp behavior as in \eqref{Ecuspbehavior} and
\ie 
\left(\Delta+{d(d-2)\over 4}\right)\cE'=\cE_{\rm reg} \,. 
\fe
Then $\cE-\cE' \in L^2(\cF)$ and 
\ie 
\left(\Delta+{d(d-2)\over 4}\right)(\cE-\cE')=0\,,
\fe
which is not possible by the spectral decomposition \eqref{RSdecomp}, of which the eigenspectrum for $\Delta$ is bounded below by  $\frac{1}{4}$.

It is convenient to work with the symmetric Eisenstein series $E^*_s(\tau)$, correspondingly we define 
\ie 
\{f,E_s\}\equiv {\la f, E_s\ra \over \Lambda(s)}\,,\quad \{f,E_s\}=\{f,E_{1-s}\}\,,
\fe
then we can write,
\ie 
\cE(\tau)
=&\,c_1 E_{{d\over 2}}(\tau)
+{12\over d(d-2)\pi} \int_\cF {d^2\tau\over y^2} \cE_{\rm reg}(\tau) 
+
{1\over 4\pi i}
\int_{{\rm Re\,}s={1\over 2}} ds {\{ \cE_{\rm reg},E_s\} \over {(d-1)^2\over4}-{(2s-1)^2\over 4}} E^*_s(\tau)
\\
+&
\sum_{j=1}^\infty{\la \cE_{\rm reg},\n_j\ra \over \lambda_j +{d(d-2)\over 4}} \n_j(\tau)\,,
\label{CEdecomp}
\fe
Focusing on the zero Fourier mode of $\cE$, we have
\ie 
\cE_0(y)=&
c_1\varphi_{d\over 2}(y)
+{12\over d(d-2)\pi} \int_\cF {d^2\tau\over y^2} \cE_{\rm reg}(\tau) 
+{1\over 2\pi i}
\int_{{\rm Re\,}s={1\over 2}} ds {\{ \cE_{\rm reg},E_s\} \over {(d-1)^2\over4}-{(2s-1)^2\over 4}} \Lambda(s)y^s\,,
\label{E0decomp}
\fe
and similarly for the zero mode of $\cE_{\rm reg}$
\ie 
(\cE_{\rm reg})_0
={3\over \pi} \int_{\cF} {d^2\tau\over y^2} \cE_{\rm reg}(\tau)
+
{1\over 2\pi i}
\int_{{\rm Re\,}s={1\over 2}} ds \{ \cE_{\rm reg},E_s\} \Lambda(s)y^s\,.
\label{H0decomp}
\fe
We have used here that non-constant cusp forms do not have zero Fourier mode.

A major advantage of the spectral decomposition is that it isolates the potential unknowns and packages them into the coefficients, which we define as
\ie 
h(s)\equiv \{\cE_{\rm reg},E_s\}\,,\quad p_j\equiv \la \cE_{\rm reg},\n_j\ra~{\rm for}~j\geq 1\,,
\label{overlap}
\fe
which completely determines $\cE(\tau)$. We summarize the properties of $h(s)$ below:
\begin{enumerate}
    \item $h(s)=h(1-s)$,
    \item $h(s)\in \mR$ for $s\in \mR$,
    \item $h({1\over 2})=0$,
    \item $h(0)={6\over \pi} \int_{\cF} {d^2\tau\over y^2} \cE_{\rm reg}(\tau)$,
    \item $h(s)\Lambda(s)$ is meromorphic  for $s\in \mC$ with a single  simple pole at $s=0$,
    \item $h({d\over 2})=
(d-1)
\left (
{c_2\over \Lambda({d-1\over 2})}
-
{c_1\over \Lambda({d\over 2})}
\right )
$.
\end{enumerate}
The first three properties above follow from the definition of the overlap $\{\cE_{\rm reg},E_s\}$ and that $\cE_{\rm reg}$ is square integrable. In general the spectral overlap $\{f,E_s\}$ for modular function $f\in L^2(\cF)$ has a meromorphic analytic continuation to $s\in \mC$. Here for $\cE_{\rm reg}$, we will see that $h(s)\Lambda(s)$ cannot have any poles except $s=0$.
To see this, let us study the behavior of \eqref{H0decomp} in the large $y$ region. We are then instructed to deform the contour to the left. This could potentially generate perturbative terms in $y$ from poles in the integrand. 
Since $\cE_{\rm reg}(\tau)$ cannot have perturbative terms in $y$ near the cusp, the integrand $h(s)\Lambda(s)$ must have the right pole structure to cancel the constant term in \eqref{H0decomp}. This comes from the pole of $h(s)\Lambda(s)$ at $s=0$. Since $\Lambda(s)$ has a simple pole at $s=0$ \eqref{LFres}, we conclude 
\ie 
h(0)={6\over \pi} \int_\cF {d^2\tau \over y^2} \cE_{\rm reg}(\tau)\,.
\fe
Furthermore, there 
cannot be  poles in $h(s)$ to the left of ${\rm Re}\,s=0$, and then by reflection, $h(s)$ cannot have poles anywhere for $s\in \mC$ except at zeroes of $\Lambda(s)$ along ${\Re s}=1/4$ (and its reflection along ${\Re s}=3/4$). 
Similarly by looking at the large $y$ behavior of \eqref{E0decomp}, and matching with the cusp behavior in \eqref{Ecuspbehavior}, we conclude 
\ie 
h\left({d\over 2}\right)=h\left({2-d\over 2}\right)=
(d-1)
\left (
{c_2\over \Lambda({d-1\over 2})}
-
{c_1\over \Lambda({d\over 2})}
\right )\,.
\fe
Therefore we can equivalently write, by contour deformation,
\ie 
\cE=
{1\over 2}\left ( c_1+ {\Lambda({d\over 2}) \over \Lambda({d-1\over2})} c_2\right) E_{d\over 2} (\tau)
+{1\over 4\pi i}
\int_{{\rm Re\,}s>{d\over 2}} ds {\{ \cE_{\rm reg},E_s\} \over{(d-1)^2\over4}-{(2s-1)^2\over 4}} E^*_s(\tau)
+
\sum_{j=1}^\infty{\la \cE_{\rm reg},\n_j\ra \over \lambda_j +{d(d-2)\over 4}} \n_j(\tau)\,.
\fe 
and for the zero Fourier mode,
\ie 
\cE_0
=&
{1\over 2}\left ( c_1+ {\Lambda({d\over 2}) \over \Lambda({d-1\over2})} c_2\right) \varphi_{d\over 2} (y)
+{1\over 4\pi i}
\int_{{\rm Re\,}s>{d\over 2}} ds {\{ \cE_{\rm reg},E_s\} \over {(d-1)^2\over4}-{(2s-1)^2\over 4}} \varphi_s(y)  \,.
\fe 

So far we have focused on the overlap coefficient function $h(s)$ in \eqref{overlap} that is necessary and sufficient to reconstruct the zero Fourier mode of $\cE(\tau)$. To fully determine $\cE(\tau)$ we also need the overlaps with the cusp forms 
 \eqref{cuspforms} which are denoted by $p_j$ in \eqref{overlap}. It is widely believed (though not proven) that the cusp forms do not have degenerate eigenvalues  \cite{deshouillers1985maass,hejhal1992topography,sarnak2012recent}. Consequently, the overlaps $p_j$ is determined by the $k=1$ Fourier mode of $\cE(\tau)$ using the orthogonality conditions for the Bessel functions $K_{i\n}(2\pi y)$  \cite{szmytkowski2010orthogonality}. Physically, 
 the $k$-th Fourier mode $\cE_k(y)$ of $\cE(\tau)$ accounts for the contributions from worldline instantons of total KK charge $k$ to the free energy (equivalently toroidal Casimir energy). From the above reasoning, we have just concluded that all instanton sectors of higher KK charges are completely determined by the $k=0$ and $k=1$ sectors! Such a general statement about CFT is of course a consequence of the powerful modular invariance for the toroidal geometry.\footnote{We note that this is essentially a restatement of similar constraints discussed in \cite{Benjamin:2021ygh,Collier:2022emf} albeit in a different physical context.} Furthermore, there are also constraints on the $k=0$ sector (``scalar  sector'') itself from modular invariance. In \cite{Benjamin:2022pnx}, a  closed modular bootstrap equation was derived for the scalar sector of a class of $d=2$ CFTs. Analogously, the $k=0$ mode $\cE_0$ \eqref{E0decomp} and equivalently the spectral overlap $h(s)$ in \eqref{overlap} satisfy a similar bootstrap equation. 
We leave a more detailed study of the spectral decomposition for the toroidal Casimir energy to the future.



\section{Casimir Energy Density in Free Scalar Theories} 
\label{sec:free}

For illustration of the general structure predicted by the EFT analysis in Section~\ref{sec:genstructureEFT} and the spectral theory analysis in Section~\ref{sec:spectraltheory}, here we consider the free CFT of a real scalar in dimension $d>2$. The computation of the toroidal Casimir energy for the free scalar was done in
\cite{Cappelli:1988vw,Kirsten:1995st} (and more recently for higher dimensional torus in \cite{Alessio:2021krn}). Below we review these results and compare with our general result \eqref{EFTres} from EFT analysis.


The scalar Casimir energy density on $\cM_d=T^2\times \mR^{d-3,1}$ is given by the following obvious sum-integral,
\ie 
E_{\rm vac}=  \sideset{}{'}\sum_{m,n \in \mZ} \int {d^{d-3} \vec p\over (2\pi)^{d-3}} {1\over 2}\omega_{m,n}(\vec p)\,,
\label{freesumintegral}
\fe
where each individual mode contributes,
\ie 
\omega_{m,n}({\vec p})=\sqrt{\vec p^2+{(m \vec k_1+n\vec k_2)^2}}\,,
\fe
and $\vec k_1$ and $\vec k_2$ are the momentum lattice basis vectors on the $T^2$,
\ie
    \vec{k}_1 = {2\pi\over L_1} \left(1, -\frac{\tau_1}{\tau_2} \right), \quad \vec{k}_2 = {2\pi\over L_1} \left(0, \frac{1}{\tau_2} \right) \,.
\label{momentumbasis}
\fe
Note that we have explicitly excluded the $m=n=0$ term in the sum. 

The sum-integral \eqref{freesumintegral} suffers from UV divergence. We regulate the non-compact momentum integral by dimensional regularization,
\ie 
E_{\rm vac}=-{1\over (L\tau_2)^{d-2}}{\pi^{d-2\over 2}\over 2} \Gamma\left(1-{d\over 2}\right)\sideset{}{'}\sum_{m,n\in\mZ} |m\tau+n|^{d-2}\,.
\label{freeE}
\fe
The residual sum over the momentum lattice is regulated by analytic continuation using the real analytic Eisenstein series defined in \eqref{Essym}. Taking into account the rescaling \eqref{Evactodensity}, we obtain the dimensionless modular invariant Casimir energy density for the free scalar,
\ie 
\cE(\tau)=  E^*_{2-d\over 2}(\tau)= E^*_{d\over 2}(\tau)\,,
\fe
where in the last equality we have used the reflection symmetry \eqref{Esreflection} of the Eisenstein series.

The free scalar Casimir energy density has a simple Fourier decomposition in $\tau_1$ (see \eqref{genEis}),
\ie 
{\cal E}(\tau)={  \Lambda(d/2) }\tau_2^{d/2} +{ {\Lambda(1-d/2)} }\tau_2^{1-d/2}
+4\tau_2^{1/2}\sum_{k,n=1}^\infty  \left({k\over n}\right)^{d-1\over 2}   { \cos(2\pi kn \tau_1) K_{{d-1\over 2}}(2\pi kn\tau_2)}\,.
\label{freeEfourier}
\fe 
This makes explicit the hierarchy between perturbative and non-perturbative contributions in the limit $\tau \to i\infty$. The former comes from the zero Fourier mode and comparing to \eqref{EFTres}, we find the corresponding coefficients 
\ie 
c_1=\Lambda\left({d\over 2}\right)\,,\quad c_2=\Lambda\left({d-1\over 2}\right)\,,
\label{c1c2free}
\fe
where $\Lambda(s)$ is defined in \eqref{Lambdadef}.
Indeed, the thermal one-point function of stress tensor of a free real scalar (see for example \cite{Iliesiu:2018fao}) takes the form \eqref{Tmn1PF} with
\ie 
b_T
=-  {2\zeta(d)\over d {\rm vol}(S^{d-1})}=
-{1\over d}\, \Lambda\left({d\over 2}\right)\,,
\label{btfree}
\fe
where we have used ${\rm vol}(S^{d-1})={2\pi^{d/ 2}\over \Gamma({d/2})}$. Since the gapless sector of a free scalar upon $S^1$ reduction is obviously a free scalar in one lower dimension, we see \eqref{btfree} 
is in agreement with \eqref{c1c2free} via the general relations \eqref{c1rel} and \eqref{c2rel}.

The non-perturbative contributions to the toroidal Casimir energy for the free scalar come entirely from the nonzero Fourier modes in \eqref{freeEfourier}. As explained in Section~\ref{sec:worldlineinstanton}, they are associated with worldline instantons of massive particles in the KK reduced theory. In the free scalar theory, such massive particles are in one-to-one correspondence with the $k$-th KK modes ($k\neq 0$), whose (dimensionless) mass and KK-charge satisfy $M=Q=k$. The worldline instanton from the $k$-th KK mode going $n$-times around the compact $S^1_L$ is weighted by $e^{2\pi i kn \tau}$ from its worldline action (see \eqref{SMQ}). 
Together with fluctuations thereof, they account for the nonzero Fourier modes in \eqref{freeEfourier}.

Finally, in terms of the spectral decomposition discussed in Section~\ref{sec:spectraltheory}, the free scalar toroidal Casimir energy is the special case where the only nonzero term on the RHS in \eqref{CEdecomp} is the first term. Equivalently $\cE_{\rm reg}=0$ (see \eqref{diffreg}) as a consequence of the Laplace-type differential equation \eqref{EisenLaplace} satisfied by the Eisenstein series $E^*_{d\over 2}$.

\section{Casimir Energy  Density in Interacting CFTs}
\label{sec:interacting}

The power of our general results from the previous sections lies in that it applies to general CFTs. While it is very difficult to compute the toroidal Casimir energy for interacting CFTs in general, here we discuss examples where analytical methods are applicable. We will see how these analytical results confirm our general formula \eqref{EFTres} from EFT considerations
and we will also discuss the interesting features  brought about by nontrivial interactions.

\subsection{Critical $O(N)$ Scalar CFT}
\label{sec:ON}

The first interacting CFT we consider is the critical $O(N)$ CFT in $d=3$ dimensions. This is perhaps the most widely studied class of CFTs in $d>2$. It models the second order phase transitions of three-dimensional statistical models including the uniaxial magnet ($N=1$), the XY magnet ($N=2$), the Heisenberg magnet ($N=3$) and the spherical model ($N=\infty$), as well as the quantum critical points of two-dimensional materials (see \cite{Henriksson:2022rnm} for a recent review). In the large $N$ limit, the singlet sector of the $O(N)$ CFT is conjectured to have a  holographic dual \cite{Maldacena:1997re,
Gubser:1998bc,Witten:1998qj} described by the Vasiliev's higher-spin gauge theory on AdS$_4$ \cite{Vasiliev:1990en,Vasiliev:1992av,Vasiliev:1995dn} subject to suitable boundary conditions \cite{Klebanov:2002ja,Giombi:2009wh}.\footnote{See also \cite{Aharony:2020omh,Aharony:2022feg} for recent works on this holographic duality beyond the leading large $N$ limit.} 

To describe the $O(N)$ CFT, we start with a scalar field theory defined by the following 
$O(N)$ symmetric action on flat space $\mR^3$,\footnote{We refer the readers to \cite{Moshe:2003xn} for further details on the $O(N)$ model and its large $N$ limit.}
\ie
S[\phi] = \int d^3 x \left[ \frac{1}{2} (\partial_{\mu} \phi_i)^2 +\frac{1}{2} r \phi_i\phi^i + \frac{1}{4!} \frac{u}{N} (\phi_i\phi^i)^2 \right]\,,
\label{ONact}
\fe
where $\phi_i$ with $i=1,2,\dots N$ denotes a scalar field in the vector representation of $O(N)$ and $u,r$ are the coupling constants. The quartic interaction is relevant in $d=3$ and triggers an renormalization group (RG) flow from the free $O(N)$ symmetric theory to a nontrivial fixed point in the infra-red (IR) described by the critical $O(N)$ CFT. Near the critical point, the mass term is relevant and thus needs to be tuned $r\to r_c$ to ensure a vanishing mass gap and furthermore we take $u\to \infty$ to reach the CFT.

Despite the explicit UV Lagrangian, there is no small parameter in \eqref{ONact} and thus it is not useful in practice to extract CFT data in general. Nonetheless, in the large $N$ limit, as is suggested by the form of \eqref{ONact} with $u$ fixed,  we can make use of the $1/N$ expansion.

Here the standard procedure starts with the Hubbard-Stratonovich transformation to replace the quartic interaction term by introducing an auxiliary scalar field $\lambda$,
\ie
S[\phi,\lambda] = \int d^3 x \left[ \frac{1}{2} (\partial_{\mu} \phi_i)^2 +\frac{1}{2}  r \phi_i\phi^i + \frac{6N \lambda^2}{u} + i\lambda \phi_i\phi^i\right]\,,
\fe
and path integrate over real $\lambda$.
Since $\phi$ appears quadratically in $S[\phi,\lambda]$,  we can integrate it out and obtain an action for $\lambda$,
\ie
S[\lambda] = \frac{N}{2} \Tr \ln (-\partial^2 + r + 2i \lambda) +\frac{6 N \lambda^2}{u}\,.
\fe
In the large $N$ limit, from the saddle point approximation, we obtain the following equation for $\lambda$,
\ie
\int \frac{d^3 k}{(2\pi)^3} \frac{1}{k^2 + r + 2i\lambda} = \frac{12 i\lambda}{u}\,.
\label{R3gapeqn}
\fe
This is known as the gap equation because it determines the mass gap $m_{\mR^3}^2=r+2i\lambda$.\footnote{As we will see, for the critical $O(N)$ model on $\mR\times T^2$, the saddle point of $\lambda$ is imaginary and the mass gap $m_{\mR^3}^2>0$.} The critical coupling is determined by 
\ie  
r_c=-{u\over 6}  \int {d^3k\over (2\pi)^3} {1\over k^2}\,.
\fe 
It is convenient to work with the dimensional regularization so that $r_c=0$. Therefore to study the large $N$ $O(N)$ CFT, it suffices to focus on the following action 
\ie
S[\phi,\lambda] = \int d^3 x \left[ \frac{1}{2} (\partial_{\mu} \phi_i)^2   + i\lambda \phi_i\phi^i\right]\,,
\label{ONactsim}
\fe 
where we have sent $u\to \infty$ to decouple the irrelevant deformation $\lambda^2$. Similarly the action for $\lambda$ 
simplifies to,
\ie  
S[\lambda] =   \frac{N}{2} \Tr \ln (-\partial^2   + 2i \lambda) \,. 
\label{ONactsim2}
\fe
On $\mR^3$,  the CFT is clearly described by the $\lambda=0$ saddle point, similarly for other conformally flat spacetime manifolds such as $S^3$.




\subsubsection{Casimir Energy from Gap Equation}

We now study the $O(N)$ CFT on $\mR\times T^2$ to determine the toroidal Casimir energy (see also previous work \cite{PhysRevB.94.085134}). This boils down to solving the path integral for the action \eqref{ONactsim} (equivalently \eqref{ONactsim2}) on $\mR\times T^2$. The saddle point (gap) equation for $\lambda$ becomes, 
\ie
0= \frac{1}{A} \sum_{\vec{k}} \int \frac{d\omega}{2\pi} { 1\over \omega^2 + \vec{k}^2 + 2i\lambda}
\,,
\fe
where $A$ is the area of $T^2$ and $\vec k$ is a lattice vector with the basis vectors in \eqref{momentumbasis}. We define
\ie  
2i\lambda\equiv \left(2\pi \over L_1\right)^2 \Delta^2 
\fe 
and integrate over $\omega$ to obtain,
\ie 
0=
 \sum_{m,n\in \mZ} \frac{1}{\sqrt{ |m\tau+n|^2 +  \tau_2^2\Delta^2}}
 \label{gapeqnbr}
\fe 
The gap equation (after regularization) will determine $\Delta$ as a function of the complex moduli $\tau$. 
Physically, $2\pi \Delta$ is the induced dimensionless mass of the $\phi^i$ fields due to  the nontrivial geometry, generalizing the thermal mass in the case of $ S^1_\B \times \mR^2$. 
Equivalently, it determines the one-point function of the $O(N)$ invariant operator $\phi_i\phi^i$ on $\mR \times T^2$.
In the limit the $T^2$ area $A\to \infty$, the induced mass $m^2=2i\lambda$ vanishes   which is consistent with vanishing gap in the CFT on $\mR^3$.

The large $N$ Casimir energy is then given by the following mode sum as in the theory of $N$ free scalars of mass $\Delta$ (see around \eqref{freesumintegral}),
 \ie 
 E_{\rm vac}= {N\over 2} {2\pi \over L_1\tau_2} \sum_{m,n \in \mZ}  \sqrt{|m\tau+n|^2+\tau_2^2\Delta^2}\,.
 \fe 
 Equivalently, this follows from evaluating the CFT partition function on $S^1_\B \times T^2$ 
\ie  
  E_{\rm vac}= -\lim_{\B \to \infty} {1\over \B} \log Z={N\over 2} \sum_{\vec k} \int {d\omega\over 2\pi } \log ( \omega^2+\vec k^2+2i\lambda)  \,,
  \fe
after integrating over $\omega$ and dropping extensive terms on $T^2$ (which can be absorbed by the $d=3$ cosmological constant counterterm).
 
 After a rescaling (see \eqref{Evactodensity}), we obtain the dimensionless Casimir energy density,
\ie  
 \cE(\tau)= 
 - {\pi N} \tau_2^{-{1\over 2}} \sum_{m,n\in\mZ} \sqrt{|m\tau+n|^2+\tau_2^2\Delta(\tau)^2}\,.
 \label{Ebr}
 \fe
To regularize the momentum lattice sum that appears in the gap equation \eqref{gapeqnbr} and the Casimir energy \eqref{Ebr}, we define the following generalized Eisenstein series,
\ie  
G_{s}(\tau,\Delta)={1\over 2}\pi^{-s} \Gamma(s)\sum_{m,n\in \mZ}{\tau_2^s\over (|m\tau+n|^2+\tau_2^2\Delta(\tau)^2)^{s}}.
\label{Gsdef}
\fe 
The real analytic Eisenstein series defined in \eqref{Essym} is recovered as a special case at $\Delta=0$,
\ie 
E_s^*(\tau)=\lim_{\Delta \to 0} \left( G_s(\tau,\Delta)-{1\over 2} { \Gamma(s)\over (\pi\tau_2 \Delta^{2})^s}   \right)\,.  
\label{GftoEis}
\fe
More generally, $G_s(\tau,\Delta)$ is obviously modular invariant if $\tau_2\Delta(\tau)^2$ is modular invariant (e.g. a constant), since the above can be written as a sum of Poincar\'e series,
 \ie  
G_{s}(\tau,\Delta)=\pi^{-s} \Gamma(s) \sum_{k=1}^\infty  \sum_{\C \in \Gamma_\infty\backslash PSL(2,\mZ) }{{\rm Im} (\C \tau)^{s}\over \left ( k^2+ {\rm Im} (\C \tau) \tau_2\Delta( \tau)^2 \right)^{s}} + {1\over 2} { \Gamma(s)\over (\pi\tau_2 \Delta^{2})^s}.
\fe 
Here for the application to the CFT on $\mR\times T^2$, as a consequence of large diffeomorphisms on $T^2$, we see indeed that the induced mass normalized by the $T^2$ area ${2i\lambda} A= (2\pi)^2 \tau_2 \Delta^2$ is modular invariant and therefore so is $G_s(\tau,\Delta)$.

Similar to the  Eisenstein series $E_s(\tau)$, the function $G_s(\tau,\Delta)$ admits an analytic continuation in $s$
that is finite at $s=-1/2$ and $s=1/2$ which will be relevant for the gap equation and the Casimir energy respectively. We note that the function $G_s(\tau,\Delta)$ is a slight modification of the function $g_s(\Delta,\tau)$ introduced in \cite{PhysRevB.94.085134} which they used to analyze the spectrum on $T^2$ numerically. There the analytic continuation in $s$ is provided by an integral formula. Here we will provide an alternative but equivalent formula in the form of Poincar\'e series (more precisely sum of them) which makes the modular invariance manifest,\footnote{The special case of the modular invariant function $G_s(\tau,\Delta)$ when $\tau_2\Delta^2$ is a constant has also recently shown up in a different context \cite{Dorigoni:2022cua}.}
\ie 
G_s(\tau,\Delta)=&     \sideset{}{'}\sum_{m,n\in \mZ}  \left (\frac{\tau_2 \Delta}{|n+m\tau|}\right)^{1-s} K_{1-s}(2\pi \Delta |n+m\tau|)  +{\Gamma(s-1)\over 2} \left(\pi \tau_2 \Delta^2\right)^{1-s}   \,,
\label{regGf}
\fe 
and is well defined for $\Delta$ away from the branch cut $\Delta < 0$.
The details of the derivation and the comparison to the formulae in \cite{PhysRevB.94.085134} are given in Appendix~\ref{app:Gfunc}.  

In terms of the regulated generalized Eisenstein series $G_s(\tau,\Delta)$ in \eqref{regGf}, the gap equation becomes\footnote{In other words, $\Delta$ is the fixed point of the $\tau$-dependent lattice sum.}
 \ie  
0= G_{1\over 2}(\tau,\Delta)
 \quad \Leftrightarrow \quad
\sideset{}{'}\sum_{m,n\in \mZ} \frac{1}{ |n+m\tau|} e^{-2\pi \Delta |n+m\tau|}  =2\pi \Delta \,,
\label{gapeqnON}
 \fe
 and the Casimir energy is
 \ie  
\cE(\tau,\Delta)={N} G_{-{1\over2}}(\tau,\Delta)= \frac{N \tau_2^{\frac{3}{2}}}{4\pi}  \sideset{}{'}\sum_{m,n\in \mZ} \frac{1+2\pi \Delta |n+m\tau|}{|n+m\tau|^3} e^{-2\pi \Delta|n+m\tau|} +\frac{2 N \pi^2 \Delta^3 \tau_2^{\frac{3}{2}}}{3}\,.
\label{EON}
 \fe
Note the close resemblance to the corresponding expression for the free scalar \eqref{freeE}.  The toroidal Casimir energy for the $O(N)$ CFT is obtained from solving the gap equation \eqref{gapeqnON} for the induced mass $\Delta(\tau)$ and then plugging it in \eqref{EON}.
The gap equation says that the CFT Casimir energy extremizes $\cE(\tau,\Delta)$ with respect to $\Delta^2$. The induced mass squared $\Delta^2$ must be non-negative to avoid an instability on $T^2$ (and thus divergent partition function on $T^3$). For our choice of branch cut in \eqref{regGf}, this means the saddle point has $\Delta(\tau)\geq 0$, which is clearly consistent with \eqref{gapeqnON}.
In fact, the  saddle point is a local minimum of $\cE(\tau,\Delta)$, since
\ie  
{\pa^2 \cE(\tau,\Delta) \over \pa (\Delta^2)^2}=
{N} \pi^2 \tau_2^2 G_{{3\over2}}(\tau,\Delta)= \frac{N \tau_2^{\frac{3}{2}}}{32\pi^2 \Delta}   \sum_{m,n\in \mZ}   e^{-2\pi \Delta|n+m\tau|}  \,,
\fe
which is positive for $\Delta \geq 0$.

\begin{figure}[!htb]
\begin{subfigure}{0.5\textwidth}
\scalebox{1}{\includegraphics[width=0.8\textwidth]{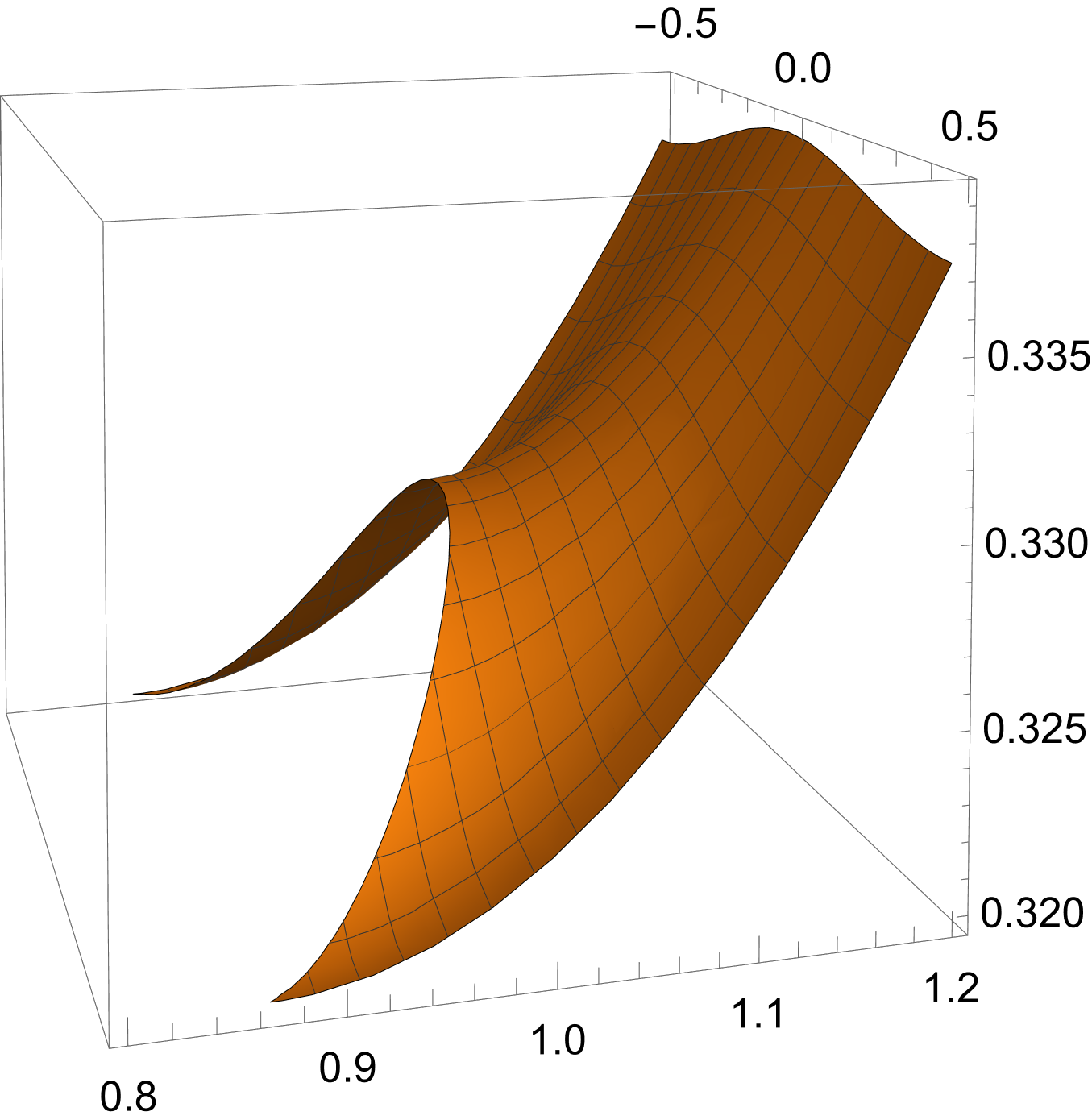}\put(-10,150){$\frac{\cE(\tau)}{N}$}\put(-190,5){$\tau_2$}\put(-100,190){$\tau_1$}}
\end{subfigure}
\begin{subfigure}{0.5\textwidth}
\scalebox{1}{\includegraphics[width=0.8\textwidth]{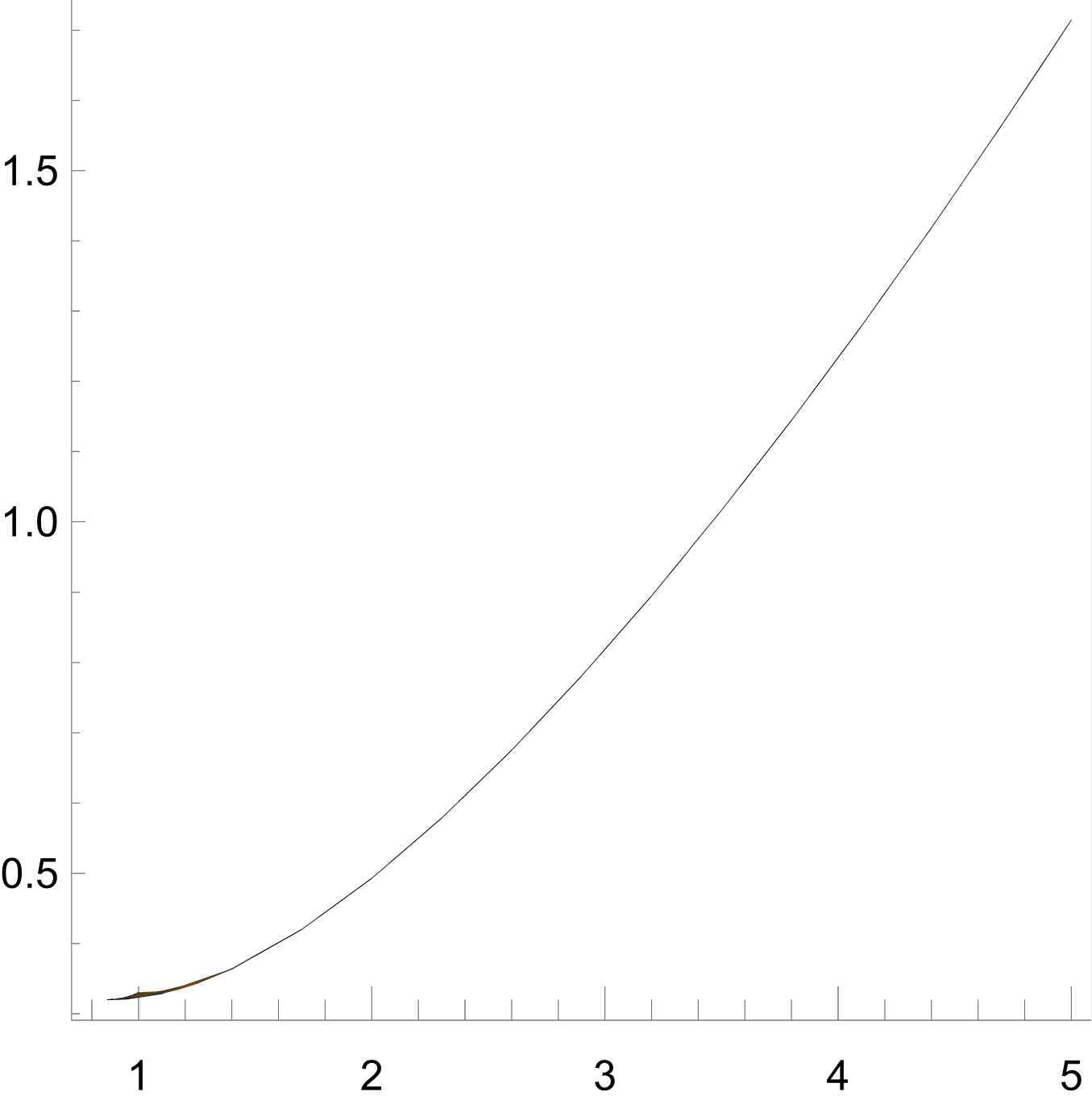}\put(-200,190){$\frac{\cE(\tau)}{N}$}\put(10,0){$\tau_2$}}
\end{subfigure}
\caption{Toroidal Casimir energy of the $O(N)$ scalar CFT as a function of $\tau$, computed in the standard $PSL(2,\mathbb{Z})$ fundamental domain. In the left figure, we zoom into the region around $\tau_2=1$ to show the $\tau_1$ dependence of the Casimir energy, which is more prominent at smaller $\tau_2$. In the right figure, we project the 3D plot along the $-\tau_1$ direction to highlight the suppression of $\tau_1$ independence in this fundamental domain. $\cE(\tau)$ grows at large $\tau_2$ as $\tau_2^{3/2}$, which becomes dominant as early as $\tau_2 \sim 1$ in the $O(N)$ scalar CFT. The $\tau_1$ dependence is negligible in this domain, except for minor dependence near the boundary at $|\tau|=1$, which is mostly contributed by the lowest nonzero KK mode. }
\label{fig:ONenergy}
\end{figure}

\begin{figure}[!htb]
\scalebox{1}{\includegraphics[width=0.8\textwidth]{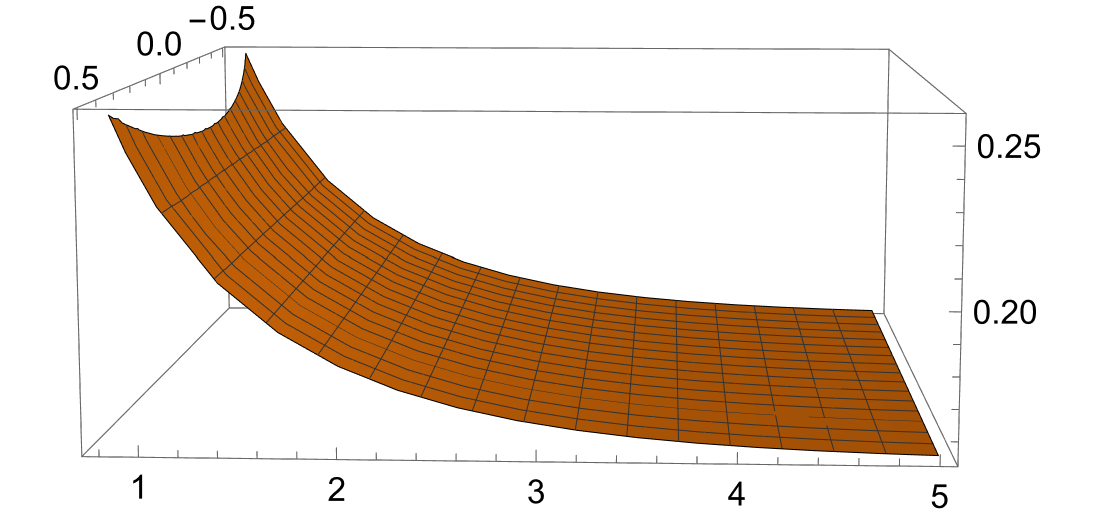}\put(-10,150){$\Delta(\tau)$}\put(-370,5){$\tau_2$}\put(-370,140){$\tau_1$}}
\centering
\caption{Rescaled induced mass $\Delta$ of the $O(N)$ scalar CFT as a function of $\tau$, computed in the standard $PSL(2,\mathbb{Z})$ fundamental domain. The induced mass hardly depends on $\tau_1$ in this $PSL(2,\mZ)$ frame, and approaches a constant at large $\tau_2$. }
\label{fig:ONgap}
\end{figure}

   
We solve the gap equation \eqref{gapeqnON}  numerically and then evaluate the toroidal Casimir energy \eqref{EON}.\footnote{For the numerical evaluation, we find another form \eqref{bessel_sum} for $G_s(\tau,\Delta)$ to be more useful, as the sum of Bessel functions converges very fast on the standard fundamental domain.} In Figure~\ref{fig:ONenergy}, we present a plot of $\cE(\tau)$ for the $O(N)$ CFT in the standard $PSL(2,\mZ)$ fundamental domain. We see the growth behavior at large $\tau_2$ as $\tau_2^{\frac{3}{2}}$, and the $\tau_1$ dependence is almost negligible in this region. In Figure~\ref{fig:ONgap}, we present a plot of $\Delta(\tau)$ in the standard $PSL(2,\mZ)$ fundamental domain. The induced mass tends toward a constant at large $\tau_2$. The dependence on $\tau_1$ is also very weak. In the next section, we  discuss analytic results for $\cE(\tau)$
in the thin torus (high temperature) limit, which will explain these observations.

\subsubsection{High Temperature Expansion and Worldline Instantons}

We now study the toroidal Casimir energy $\cE(\tau)$ for the $O(N)$ CFT in the thin torus $\tau \to i\infty$ limit, to compare with the universal formula \eqref{EFTres} derived from high temperature EFT and the numerical results obtained above. For this purpose it is useful to use another form of the regulated generalized Eisenstein series \eqref{regGf}. Specifically, we have 
\ie  
\cE(\tau,\Delta)= N{\tau_2^{3\over 2}
f(\Delta)}
+
2N\tau_2^{1\over 2} \sum_{n\in \mZ} \sum_{m=1}^\infty 
{1\over m} e^{2\pi i n m \tau_1}
\sqrt{n^2+\Delta^2}
K_1(2\pi \sqrt{n^2+\Delta^2} m \tau_2)\,,
\label{EONbessel}
\fe 
where
\ie 
f(\Delta)\equiv \frac{1}{6\pi} \left(4\pi^3 \Delta ^3+6\pi \Delta  \text{Li}_2\left(e^{-2\pi \Delta }\right)+3 \text{Li}_3\left(e^{-2\pi \Delta }\right)\right)\,.
\fe
The derivation is given in Appendix~\ref{app:Gfunc}. The gap equation, which locates the local minimum of $\cE(\tau,\Delta)$ in $\Delta$, becomes
\ie 
\pi \Delta+ \log (1-e^{-2\pi \Delta})
=2\sum_{n\in \mZ} \sum_{m=1}^\infty 
e^{2\pi i n m \tau_1}
K_0(2\pi \sqrt{n^2+\Delta^2} m \tau_2)\,.
\label{GEbessel}
\fe 

Let us first analyze the gap equation \eqref{GEbessel} and Casimir energy \eqref{EONbessel} at infinite $\tau_2$. From positivity and asymptotic behavior the Bessel function $K_0(x)$, it is clear that the saddle point $\Delta$ that solves \eqref{GEbessel} approaches a constant $\Delta_0\equiv \Delta(\tau=i\infty)$ which satisfies 
\ie  
\pi \Delta_0+\log (1-e^{-2\pi \Delta_0})=0\quad \Rightarrow \quad  \Delta_0 = \frac{1}{\pi} \log (\frac{1+\sqrt{5}}{2})\,.
\label{Dinf}
\fe
Consequently, the Casimir energy evaluates to
\ie  
\cE(\tau=i\infty)=
\frac{N\tau_2^{3\over 2}}{6\pi} \left(4\pi^3 \Delta_0 ^3+6\pi \Delta_0  \text{Li}_2\left(e^{-2\pi \Delta_0}\right)+3 \text{Li}_3\left(e^{-2\pi \Delta_0}\right)\right)= {2N \zeta(3)\over 5\pi} \tau_2^{3\over 2}\,.
\label{Einf}
\fe
This determines the coefficient $c_1$ in \eqref{EFTres} for the $O(N)$ CFT in the large $N$ limit to be 
\ie  
c_1^{O(N)}={2N\zeta(3)\over 5\pi} \,.
\fe
Here $2\pi \Delta_0$ coincides with the dimensionless thermal mass \cite{Chubukov_1994} on $\mathbb{R}^2 \times S^1_\B$ and $c_1$ agrees with the thermal free energy \cite{Sachdev:1993pr} via the relation \eqref{c1rel}.

A main advantage of the equations \eqref{GEbessel} and \eqref{EONbessel} is that it can be solved recursively in a large $\tau_2$ expansion. First we note that by dropping all exponentially suppressed terms in \eqref{GEbessel} and \eqref{EONbessel}, there is no perturbative (i.e. power law in $\tau_2$) correction to \eqref{Dinf} and \eqref{Einf}. Consequently we conclude that $c_2$ in \eqref{EFTres} vanishes for the $O(N)$ CFT to the leading order in the large $N$ limit,
\ie  
c_2^{O(N)}=0\,.
\fe 
This is consistent with the fact that the circle reduced theory (keeping only zero 
 KK modes), namely the $O(N)$ vector model  in $d=2$ (equivalent to the  $O(N)$ $\sigma$-model in the IR) famously has a mass gap for $N\geq 3$ 
\cite{Brezin:1975sq,Polyakov:1975rr,Brezin:1976qa,Bardeen:1976zh,Polyakov:1983tt,Wiegmann:1985jt,Hasenfratz:1990zz,Hasenfratz:1990ab}.

Given the general discussion in see Section~\ref{sec:worldlineinstanton}, we are then led to the following expansion of the solution $\Delta(\tau)$ for the induced mass,
\ie
\Delta(\tau) = \Delta_0 + \sum_{j=1}^\infty \sum_{\substack{\vec m \in \mathbb{Z}^j, \\ 
m_1 \leq m_2 \leq \cdots \leq m_j}}    e^{- 2\pi \tau_2 \sum_{i=1}^j \sqrt{m_i^2+\Delta_0^2}} 
e^{2\pi i \sum_{i=1}^j  m_i \tau_1}\Delta_{j,\vec m}(\tau_2)\,.
\fe 
The second term on the RHS above keeps track of the contributions from multiple worldline instantons in the EFT from KK reduction.
Here $j$ counts the number of worldline instantons and each individual instanton labeled by $i=1,\dots,j$
has KK charge $Q=m_i$ and mass $M=\sqrt{m_i^2+\Delta_0^2}$. 

Similarly for the Casimir energy, we have the following expansion,
\ie
\cE(\tau) = c_1\tau_2^{\frac{3}{2}}
+
N \sum_{j=1}^\infty \sum_{\substack{\vec m \in \mathbb{Z}^j, \\ 
m_1 \leq m_2 \leq \cdots \leq m_j}}   e^{- 2\pi \tau_2 \sum_{i=1}^j \sqrt{m_i^2+\Delta_0^2}} 
e^{2\pi i \sum_{i=1}^j  m_i \tau_1}\cE_{j,\vec m}(\tau_2)\,.
\label{ONEexpansion}
\fe 
Here the coefficient functions $\Delta_{j,\vec m}(\tau_2)$ and $\cE_{j,\vec m}(\tau_2)$ are perturbative in $\tau_2$ and account for fluctuations on the multi-instanton background and integrations over moduli of such configurations.  
Furthermore, parity invariance of the $O(N)$ CFT  implies that 
\ie  
\Delta_{j,\vec m}(\tau_2)=\Delta_{j,-\vec m}(\tau_2)\,,\quad \cE_{j,\vec m}(\tau_2)=\cE_{j,-\vec m}(\tau_2)\,.
\fe

One can solve for $\Delta_{j,\vec{m}}$ from \eqref{GEbessel} recursively starting from $j=1$, and then  evaluate \eqref{EONbessel} on the solution. 
Below we denote $\cK_{\nu} (z) \equiv  K_{\nu} (z) e^{z}$ and  $M_i \equiv \sqrt{m_i^2 + \Delta_0^2}$ for convenience, and list the solutions that account for the single-instanton effects,
\ie
\Delta_{1,m_1} =& \, \frac{2}{\sqrt{5} \pi} \cK_0 \left( 2\pi M_1 \tau_2 \right) 
=\frac{1}{\pi} \sqrt{\frac{1}{5 M_1 \tau_2}} \left( 1 - \frac{1}{16\pi M_1 \tau_2} + \cO(\tau_2^2) \right)\,, 
\label{mass_expand1}
\fe
the two-instanton contributions,
\ie
\Delta_{2,m_1,m_2} =& \, \frac{2\pi}{ \sqrt{5} } \Delta_{1,m_1} \Delta_{1,m_2} + \frac{2}{\sqrt{5} \pi} \cK_0 \left( 4\pi M_1 \tau_2 \right) \delta_{m_1,m_2} \\
-& \, \frac{8 \Delta_0 \tau_2}{5\pi M_1} \cK_1 \left( 2\pi M_1 \tau_2 \right) \cK_0 \left( 2\pi M_2 \tau_2 \right) + \text{distinct perm. of } \{m_i\} \\
=& -\frac{2 \Delta_0}{5\pi} \frac{1}{M_1^{\frac{3}{2}} M_2^{\frac{1}{2}} } + \frac{\delta_{m_1,m_2} }{\pi \sqrt{10 M_1} } \tau_2^{-\frac{1}{2}} + \cO(\tau_2^{-1}) + \text{distinct perm. of } \{m_i\} \,,
\label{mass_expand2}
\fe
and three-instanton contributions,
\ie
& \Delta_{3,m_1,m_2,m_3} = \, \frac{4\pi}{\sqrt{5}} \Delta_{1,m_1} \Delta_{2,m_2,m_3} - \frac{4\pi^2}{3} \Delta_{1,m_1} \Delta_{1,m_2} \Delta_{1,m_2} + \frac{2}{\sqrt{5} \pi} \cK_0 \left( 6\pi M_1 \tau_2 \right)\delta_{m_1,m_2}\delta_{m_1,m_3} \\
 +& \, \frac{2\tau_2}{\sqrt{5} M_1^3} \left[ 2\pi \tau_2 \Delta_0^2 M_1 \, \cK_0 \left( 2\pi M_1 \tau_2 \right) + (-m_1^2 + \Delta_0^2) \cK_1 \left( 2\pi M_1 \tau_2 \right) \right] \Delta_{1,m_2} \Delta_{1,m_3} \\ 
-& \, \frac{8 \Delta_0 \tau_2}{\sqrt{5}} \frac{\cK_1 \left( 4\pi M_1 \tau_2 \right)}{M_1} \delta_{m_1,m_2} \Delta_{1,m_3} - \frac{4 \Delta_0 \tau_2}{\sqrt{5}} \frac{\cK_1 \left( 2\pi M_1 \tau_2 \right)}{M_1} \Delta_{2,m_2,m_3} + \text{distinct perm. of } \{m_i\} \\
=& \, \frac{2 \Delta_0^2 \tau_2^{\frac{1}{2}} }{5\sqrt{5} \pi} \left( \frac{1}{M_1^{\frac{5}{2}} M_2^{\frac{1}{2}} M_3^{\frac{1}{2}} } +  \frac{2}{M_1^{\frac{3}{2}} M_2^{\frac{3}{2}} M_3^{\frac{1}{2}}} \right) - \frac{\sqrt{2} \Delta_0}{5\pi} \left( \frac{2}{M_1^{\frac{3}{2}} M_3^{\frac{1}{2}}} + \frac{1}{M_1^{\frac{1}{2}} M_3^{\frac{3}{2}}} \right)  \delta_{m_1,m_2} \\
+& \, \cO(\tau_2^{-\frac{1}{2}}) + \text{distinct perm. of } \{m_i\} \,.
\label{mass_expand3}
\fe
Similarly for the toroidal Casimir energy of the $O(N)$ CFT, we find the following one-instanton contributions,
\ie
\cE_{1,m_1} =& \, 2 \tau_2^{ \frac{1}{2} } M_1 \, \cK_1 \left( 2\pi M_1 \tau_2 \right)  
=\sqrt{M_1} \left(1+ \frac{3}{16\pi M_1 \tau_2} + \cO(\tau_2^{-2}) \right) \,,
\label{cas_expand1}
\fe
two-instanton contributions,
\ie
\cE_{2,m_1,m_2} =& - \sqrt{5} \pi^2 \Delta_0 \tau_2^{ \frac{3}{2} } \Delta_{1,m_1} \Delta_{1,m_2} +  \tau_2^{ \frac{1}{2} } M_1 \, \cK_1 \left( 4\pi M_1 \tau_2 \right)  \delta_{m_1,m_2} + \text{distinct perm. of } \{m_i\} \\
= & -\frac{ \tau_2^{\frac{1}{2}} \Delta_0 }{\sqrt{5 M_1 M_2}} + \frac{\sqrt{2 M_1} }{4} \delta_{m_1,m_2} + \cO(\tau_2^{-\frac{1}{2} }) + \text{distinct perm. of } \{m_i\} \,, 
\label{cas_expand2}
\fe
and three-instanton contributions,
\ie
\cE_{3,m_1,m_2,m_3} =& \frac{2\pi^2 (\sqrt{5} - 2\pi \Delta_0) }{3} \tau_2^{3/2} \Delta_{1,m_1} \Delta_{1,m_2} \Delta_{1,m_3} + \frac{2\tau_2^{ \frac{1}{2} } }{3} M_1 \, \cK_1 \left( 6\pi M_1 \tau_2 \right) \delta_{m_1,m_2}\delta_{m_1,m_3} \\
+& \, 2\tau_2^{\frac{1}{2}} \left[ \frac{1 + 2\pi^2 \tau_2^2 \Delta_0^2 }{M_1 } \, \cK_0 \left( 2\pi M_1 \tau_2 \right) - \pi \tau_2 \cK_2 \left( 2\pi M_1 \tau_2 \right) \right] \Delta_{1,m_2} \Delta_{1,m_3} \\
-& \, 4\pi \Delta_0 \tau_2^{ \frac{3}{2} } \cK_0 \left( 4\pi M_1 \tau_2 \right) \delta_{m_1,m_2} \Delta_{1,m_3} + \text{distinct perm. of } \{m_i\}  \\
=& \, \frac{2 \Delta_0^2 \tau_2}{5 } \frac{1}{ M_1^{\frac{3}{2}} M_2^{\frac{1}{2}} M_3^{\frac{1}{2}}} - \sqrt{\frac{2}{5}} \frac{\Delta_0}{ \sqrt{M_1 M_3} } \delta_{m_1,m_2} \tau_2^{\frac{1}{2}} + \cO(1) + \text{distinct perm. of } \{m_i\} \,.  
\label{cas_expand3}
\fe 


\begin{figure}[!htb]
\scalebox{.65}{\includegraphics[width=0.9\textwidth]{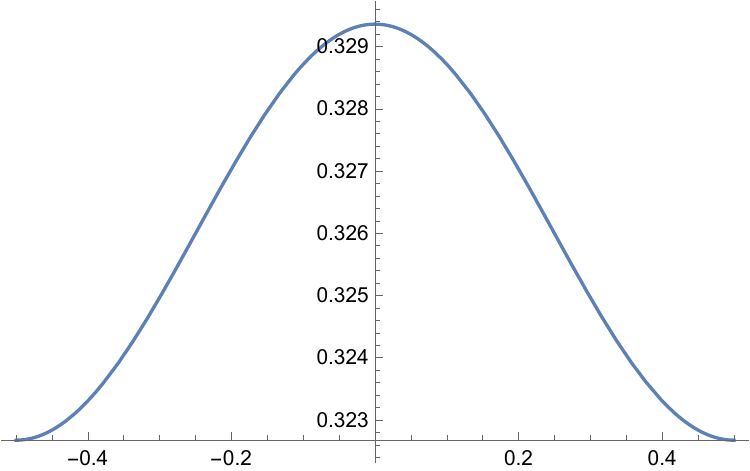}\put(-220,280){\large $\frac{\cE(\tau)}{N}$}\put(0,0){\large$\tau_1$}}
\centering
\caption{Casimir energy of $O(N)$ scalar CFT as a function of $\tau_1\in[-1/2,1/2]$ at fixed $\tau_2=1$. }
\label{fig:ONenergy_tau2section}
\end{figure}

 Because there is only one perturbative term in $\cE(\tau)$ for the $O(N)$ CFT, it is the dominant contribution on the standard fundamental domain even for not very large $\tau_2$. 
This explains why there is little $\tau_1$ dependence for $\cE(\tau)$ in this domain (see Figure~\ref{fig:ONenergy}).
Furthermore, among the instanton contributions, the leading $\tau_1$ dependence is exponentially suppressed with a larger exponent than the leading $\tau_1$ independent exponential term, due to the higher mass of instanton with KK charge. We have checked these leading instanton contributions presented above  numerically at large $\tau_2$ and they agree with the numerical solution from directly solving \eqref{gapeqnON} and then evaluating \eqref{EON} on the saddle. Specifically, we plot the  Casimir energy density $\cE(\tau)$ for the $O(N)$ CFT as a function of $\tau_1$ at $\tau_2=1$ in Figure~\ref{fig:ONenergy_tau2section}. One can compute the leading contribution to the amplitude of the first Fourier term in \eqref{ONEexpansion} 
using  \eqref{cas_expand1} which gives
$2\cE_{1,1} (\tau_2=1) e^{-2\pi \sqrt{1+\Delta_0^2}} \approx 0.0035$, while the amplitude shown in Figure~\ref{fig:ONenergy_tau2section} is about 0.0033. We see that even for $\tau_2=1$ (and it gets better for large $\tau_2$), the exponential suppression is prominent and the one-instanton coefficient already works well (about 5\% error) for estimating $\tau_1$ dependence.

\subsection{Holographic CFTs with Einstein Gravity Duals }
\label{sec:holo}

Let us now consider the case of general $d\geq 3$-dimensional large $N$ CFTs 
that are dual to Einstein gravity on AdS$_{d+1}$ at strong coupling \cite{Maldacena:1997re,
Gubser:1998bc,Witten:1998qj}. While the ground state of the CFT on $S^{d-1}$ is dual to the empty global AdS$_{d+1}$, 
 the CFT ground state on $\mR^{d-3}\times T^2$  is expected to be dual to certain
  AdS soliton solution 
\cite{Horowitz:1998ha,Myers:1999psa,Belin:2016yll} . For rectangular $T^2=S^1_{L_1}\times S^1_{L_2}$, the AdS soliton solution  is given in  \cite{Horowitz:1998ha},
 \ie 
 ds^2=-{r^2\over \ell^2} dt^2+{\ell^2 dr^2 \over r^2(1-(r_0/r)^d)} + {r^2\over \ell^2} (1-(r_0/r)^d) L_1^2 dx_1^2+{r^2 \over \ell^2} L_2^2 dx_2^2+{r^2\over \ell^2} dy_i^2
 \label{soliton}
 \fe
 which is related by a double Wick rotation to the black $d$-brane solution.
 Here $\ell$ denotes the AdS radius. The AdS radial coordinate is $r$ which ends at $r=r_0$ in the interior and approaches $r=\infty$ at the asymptotic boundary. The boundary $T^2$ coordinates are $x_1,x_2$ 
which obey the identification $x_1 \sim x_1+ 1\,,~
  x_2 \sim x_2 +1$. The remaining boundary noncompact 
 spatial directions are $y_i$ with $i=1,\dots, d-3$ and the time direction is $t$.  

The circle in the $x_1$ direction shrinks in the bulk as $r$ decreases. To avoid a singularity at $r=r_0$, this requires\footnote{Physically, to create a conical singularity requires a massive codimension-two object (e.g. worldsheet of a string in ${\rm AdS}_4$). One can check that the energy density of the AdS soliton solution \eqref{soliton} is minimized when the conical singularity is absent, so the non-singular AdS soliton represents the ground state of the dual CFT. } 
\ie
r_0={4\pi \ell^2\over d L_1}\,,
\label{solitonr0}
\fe 
so that the $x_1$ circle caps off smoothly. Meanwhile the circle in the $x_2$ direction remains non-contractible. The energy density (in the non-compact $\mR^{d-3}$ directions) for the AdS soliton is \cite{Horowitz:1998ha} 
\ie 
 E_{\rm soliton}=-{r_0^d   L_1 L_2\over 16\pi  G_N \ell^{d+1}}
 =
 -{(4\pi)^{d-1} \ell^{d-1} \over 4 d^d G_N } {L_2\over L_1^{d-1}}
  \,,
  \label{Esolitonrectangulartorus}
 \fe
 where $G_N$ denotes the Newton's constant. 

The solution of \cite{Horowitz:1998ha} for rectangular boundary $T^2$ has a straightforward generalization for general $T^2$ with complex moduli $\tau=\tau_1+i\tau_2$,
 \ie 
 ds^2=-{r^2\over \ell ^2} dt^2+{\ell ^2 dr^2 \over r^2(1-(r_0/r)^d)} + {r^2  L_1^2\over \ell^2} \left(  (1-(r_0/r)^d) (dx_1 + \tau_1 dx_2)^2+  \tau_2^2 dx_2^2 \right) + {r^2  \over \ell^2} dy_i^2\,,
 \label{solitongen}
 \fe
 where $r_0$ is determined by the same relation \eqref{solitonr0} so that the circle in the $x_1$ direction caps off smoothly in the bulk.\footnote{Topologically, the AdS soliton is a fibration of $S^1\times \mR^{d-2}$ over a two-disk (parametrized by $r$ and $x_1$). 
 }
 
The energy density for this general AdS soliton solution follows from a similar calculation,
 \ie  
 E_{\rm soliton}=-{r_0^d  A \over 16\pi  G_N \ell^{d+1}}
 =
 -{(4\pi)^{d-1} \ell^{d-1} \over 4 d^d G_N } {\tau_2^{d\over 2}\over A^{d-2\over 2}}
  \,,
  \label{Esoliton}
 \fe
 where we have used $A=L_1^2\tau_2$ as the area of the $T^2$.
 
For a fixed boundary geometry $T^2\times \mR^{d-3}$ (i.e. fixed complex moduli $\tau_1$), clearly there are multiple AdS soliton solutions related by $PSL(2,\mZ)$ transformations, corresponding to filling in different one cycles on the $T^2$ in the bulk. The CFT ground state then naturally corresponds to the AdS soliton that has the minimal energy density. From \eqref{Esoliton}, this comes from minimizing the length $L_1$ of a non-contractible cycle on the boundary $T^2$ that becomes contractible in the bulk. The minimal length of a non-contractible cycle on a compact manifold $\Sigma$ is known as the systole and denoted by ${\rm sys} (\Sigma)$. Therefore 
 the solution \eqref{solitongen} is the bulk dual for the CFT ground state when 
the $x_1$ direction is along the shortest geodesic on the $T^2$.
The rescaled dimensionless Casimir energy density (see \eqref{Evactodensity}) is determined by the systole on $T^2$,
\ie  
{\cal E}(\tau) = {(4\pi)^{d-1} \ell^{d-1} \over 4 d^d G_N } {A^{d\over2}\over {\rm sys} (T^2)^d}\,.
\label{largeNE}
\fe  
which is manifestly modular invariant since ${\rm sys} (T^2)$ is invariant.

In the limit of large $\tau_2$, we see \eqref{largeNE} matches onto  the universal formula \eqref{EFTres} from EFT analysis with
\ie  
c_1^{\rm }={(4\pi)^{d-1} \ell^{d-1} \over 4 d^d G_N } \,,\quad c_2=0\,, 
\label{c1largeN}
\fe
where $c_1$ is proportional to the thermal free energy \eqref{fbeta} (see \eqref{c1rel}), which is natural from the bulk due to the relation between the black brane (which dominates the canonical ensemble) and the AdS soliton by double Wick rotations \cite{Horowitz:1998ha}. 
Furthermore, the vanishing $c_2$ implies that upon circle reduction, the boundary CFT is completely gapped in the large $N$ limit. Finally, there are no non-perturbative contributions in the standard fundamental domain, which is related to the non-smooth feature of \eqref{largeNE}.\footnote{See \cite{Shaghoulian:2016xbx} for another perspective on the restricted dependence \eqref{largeNE} of $\cE(\tau)$ on the boundary geometry.}

The AdS/CFT dictionary determines the bulk parameters $\ell,G_N$ in terms of the boundary CFT data. 
While the detailed relations depend on the specific AdS/CFT dual pairs, the general property is that in the large $N$ limit, the $1/N$ corrections in the CFT correspond to higher derivative interactions in the bulk quantum gravity. When there is another tunable parameter in the large $N$ CFT, such as the marginal coupling $g_{\rm YM}$ in the $\cN=4$ super-Yang-Mills 
 theory (SYM), one may further take the 't Hooft limit where the 't Hooft coupling $\lambda$ (e.g. $\lambda=N g_{\rm YM}^2 $ in the SYM) is fixed as $N\to \infty$. This is possible for CFTs with string theory duals, where non-planar (higher-genus) and nonperturbative (in string coupling) effects are suppressed in the 't Hooft limit. In Table~\ref{tab:Holoc1}, we gather the relevant AdS/CFT dictionary for the following well-known AdS/CFT dual pairs,  in $d=3$ between type IIA string theory on $AdS_4\times  \mathbb{CP}^3$ and the $U(N)_k\times U(N)_{-k}$ ABJM theory \cite{Aharony:2008ug}, in $d=4$ between type IIB string theory on $AdS_5\times S^5$ and the $\cN=4$ $SU(N)$ super-Yang-Mills theory \cite{Maldacena:1997re}, in $d=5$ between type I' string theory (type IIA with O8 orientifold) on the warped background $AdS_6\times_{\rm w} {\rm HS}^4$ and the 5d $\cN=1$ rank $N$ Seiberg theories with $N_f<8$ fundamental flavors \cite{Ferrara:1998gv,Brandhuber:1999np}, and in $d=6$ between M-theory on $AdS_7\times S^4$ and the 6d $\cN=(2,0)$ theories \cite{Maldacena:1997re}. Also included in Table~\ref{tab:Holoc1} is the coefficient $c_1$ that determines the toroidal Casimir energy for these large $N$ CFTs to the leading order in the large $N$ limit via \eqref{largeNE} and \eqref{c1largeN} which is valid at strong coupling in the CFT.\footnote{\label{fn:ss}These AdS/CFT dual pairs as defined are fermionic and thus require a spin structure on the boundary manifold that extends into the bulk. Here we have implicitly summed over the spin structures on 
 $\cM_d=T^2\times \mR^{d-2}$ to produce a bosonic CFT on the boundary so that the general modular-invariant large $N$ formula \eqref{largeNE} still applies. If we choose to work directly in the fermionic CFT with a bounding spin structure $\rho$, then ${\rm sys}(T^2)$ in \eqref{largeNE} needs to be replaced by the shortest geodesic length on $T^2$ along an anti-periodic cycle for the fermion
 (see general comments in Section~\ref{sec:discussion} on toroidal Casimir energy in fermionic CFT).} 

 At subleading orders in $1\over N$, there are corrections to \eqref{largeNE} and \eqref{c1largeN} coming from higher derivative interactions in the bulk. In type IIA/IIB string theory and M-theory, the leading higher derivative term takes the schematic form $\cR^4$ which is a quartic term in the Riemann curvature tensor with a unique supersymmetric completion \cite{Green:1997tv}. 
 The $\cR^4$ term modifies the AdS soliton solution and also the energy density.
 Using the AdS/CFT dictionary (see Table~\ref{tab:Holoc1}), this leads to a correction to $c_1$ in \eqref{c1largeN} at $\cO(N^{1\over 2})$ for the 3d ABJM and the 4d $\cN=4$ SYM, and at $\cO(N)$ for the 5d $\cN=1$ Seiberg theories\footnote{The calculation in the AdS$_6$ dual of the 5d Seiberg theories will be more subtle than the other cases because of  divergences in the 10d background at the locus of the flavor D8 branes and the O8 orientifold plane (see for example \cite{Chang:2017mxc} where these divergences cancel for certain physical observables in the 5d CFT).} and the 6d $\cN=(2,0)$ theories. 
 
 The precise coefficient for this correction has been computed explicitly for the $\cN=4$ SYM in \cite{Gubser:1998nz} in the 't Hooft limit with large 't Hooft coupling $\lambda=Ng^2_{\rm YM}$,\footnote{More precisely, \cite{Gubser:1998nz} analyzed the corrections to the black brane solution in AdS$_5$ and changes to the thermal free energy due to the $\cR^4$ interaction. The corrections for the AdS soliton is then obtained from a double Wick rotation. It would be interesting to generalize this analysis to other holographic duals in string or M-theory.}
 
 \ie  
c_1^{\rm{SYM}}={\pi^2\over 8}N^2 
\left (1+{15 \zeta(3)\over 8} \lambda^{-{3\over 2}} +\cO (\lambda^{-{5\over 2}})\right) +\cO(N^0)\,,
\label{SYMc1thooft}
 \fe
 where the first term on the RHS captures all the planar contributions and the further subleading terms in $1\over \lambda$ come from bulk interactions at even 
higher derivative orders such as $D^4\cR^4$ \cite{Green:1999pu}. The non-planar contributions are contained in the second term on the RHS.
 Using the fact that the $\cR^4$ interaction depends on the type IIB axion-dilaton, equivalently the complexified Yang-Mills coupling $\tau_{\rm YM}\equiv {4\pi i\over g_{\rm YM}^2}+{\theta\over 2\pi}$, through the real analytic Eisenstein series $E_{3\over 2}(\tau_{\rm YM})$ \cite{Green:1997tv}, the large $N$ expansion of $c_1^{\rm SYM}$ at fixed $\tau_{\rm YM}$ is given by,
 \ie  
c_1^{\rm{SYM}}={\pi^2\over 8}N^2+ {15 \zeta(3)\sqrt{\pi}\over 512} \sqrt{N} E_{3\over 2}(\tau_{\rm YM}) + \cO(N^0)\,,
\label{SYMc1largeN}
 \fe
 which also makes manifest the $SL(2,\mZ)$ duality invariance in $\tau_{\rm YM}$ of the type IIB string theory and the $\cN=4$ SYM.

\begin{table}[!htb]
    \centering
    \renewcommand{\arraystretch}{1.28}
    \begin{tabular}{|c|c|c|c|c|}
        \hline  Dim & ${\ell^{d-1}/ G_N}$ & $\ell $ &  $\lambda$  & $c_1$ \\ \hline \hline 
         $d=3$ & ${ 2\sqrt{2} \over 3} k^{1/ 2}N^{3/2}  $
         &
$(\pi^2/2)^{1/6} N^{1/6}k^{1/6}\ell_s$
&  ${N\over k}$ & ${8\sqrt{2}\pi^2\over 81} k^{1/2} N^{3/2} $
\\\hline 
         $d=4$ & ${ 2 \over \pi } N^2 $
         &
        
${g_{\rm YM}^{1/ 2} N^{1/ 4} } \ell_s$
&  ${Ng_{\rm YM}^2}$ & $ {\pi^2\over 8}N^2 $
\\\hline 
         $d=5$ & $ {8\sqrt{2}\over 15\pi} \sqrt{8-N_f}N^{5/2} $
         &
        
$ (18\pi^2)^{1/4} (8-N_f)^{-1/4} N^{1/4} \ell_s$
&   \cellcolor{black!25}  & $ {512\sqrt{2} \pi^3\over 46875}{\sqrt{8-N_f}N^{5/2}} $
\\\hline 
         $d=6$ & $ {16\over 3\pi^2}  N^3 $
         &
        
$  (8\pi)^{1/3} N^{1/3} \ell_{11}$
&  \cellcolor{black!25}  & ${64\pi^3\over 2187} N^3$\\\hline 
    \end{tabular}
    \caption{The AdS/CFT dictionary involving the Newton's constant $G_N$ and the AdS scale $\ell$ on AdS$_{d+1}$ 
    for the $d=3$ ABJM CFT, the $d=4$ $\cN=4$ $SU(N)$  SYM, the $d=5$ rank $N$ Seiberg theories and the $d=6$ $SU(N)$ $\cN=(2,0)$ theories. Here $\ell_s$ is the string length and $\ell_{11}$ is the Planck length in M-theory. The 't Hooft couplings $\lambda$ are listed for the $d=3,4$ examples where an 't Hooft limit exists.}
    \label{tab:Holoc1}
\end{table}

Contrary to $c_1^{\rm SYM}$ which receives a $\cO(N^{1/2})$ correction, the coefficient $c_2^{\rm SYM}$ in \eqref{c1largeN} stays zero at $\cO(N^{1/2})$.\footnote{See Section 3 of \cite{Gubser:1998nz} for the modified black brane solution after taking into account the $\cR^4$ interaction. The modified AdS soliton (with rectangular $T^2$ on the boundary) is obtained from double Wick rotation and compactifying one of the remaining noncompact boundary directions. Using the results there, it is easy to see that the energy density for the AdS soliton still depend on $L_2,L_1$ via $L_2/L_1^{d-1}$ as in \eqref{Esolitonrectangulartorus} (i.e. extensive in $L_2$). Consequently $c_2^{\rm SYM}=0$ to this order in the large $N$ expansion (otherwise it would lead to non-extensive $L_2$ dependence in $E_{\rm soliton}$).} More generally, we expect $c_2^{\rm SYM}=0$ to all orders in $1\over N$ (at nonzero $g_{\rm YM}$). This is because the SYM reduced on a circle with thermal boundary condition\footnote{As explained in footnote~\ref{fn:ss}, we are summing over spin structures $\rho$ on $T^2$. Out of the four spin structures, three of them are bounding (even) and the remaining one is nonbounding (odd). Only the odd spin structure (periodic boundary condition for the fermions along all cycles) is compatible with supersymmetry (SUSY) and the SUSY algebra ensures the ground state energy in this sector vanishes exactly. Therefore the actual ground state of the full theory is defined in an even spin structure (which has negative energy). For a similar reason (SUSY is preserved by periodic boundary conditions for fermions), in the thin torus limit, out of the three even spin structures, the one with anti-periodic (thermal) boundary condition along the small cycle dominates. } 
for the fermions is described at low energy by the pure $d=3$ Yang-Mills theory, which confines and has a mass gap \cite{munster1981strong,Karabali:1997wk,Witten:1998zw}.
Nonetheless, we expect the Casimir energy $\cE^{\rm SYM}$ to behave drastically different at $\cO(N^0)$ where one-loop effects in gravity enter. In particular, the one-loop determinants for the bulk fields will depend nontrivially on the complex moduli $\tau$ of the torus, producing non-perturbative terms in $\tau_2$ that modifies \eqref{largeNE} at $\cO(N^0)$. 

If we work in the 't Hooft limit and focus on the planar contribution to $\cE^{\rm SYM}$,  it suffices to consider tree-level string theory in the bulk.
In this case, we expect the $\tau$ dependence in \eqref{largeNE} for the large $N$ Casimir energy to persist to all orders in $1\over \lambda$,\footnote{We have already discussed why $c_2^{\rm SYM}=0$ from the field theory side (at nonzero coupling). This can also be seen concretely from the bulk string theory in the 't Hooft limit. At tree level, the translation symmetry along $T^2\times \mR^{d-2}$ is preserved in the (deformed) AdS soliton and the metric only depends on the size of the contractible direction (e.g. $x_1$ direction in \eqref{solitongen}) through the regularity condition. Therefore $E_{\rm soliton}^{\rm SYM}$ is proportional to $A$ and then by dilatation symmetry its $T^2$ dependence (after rescaled by \eqref{Evactodensity}) takes the same form as in \eqref{largeNE}. }
\ie  
\lim_{\stackrel{N\to \infty}{ \lambda\,{\rm fixed}}} {1\over N^2}\cE^{\rm SYM}={\pi^2\over 8} 
f^{\rm SYM}(\lambda){A^2\over {\rm sys} (T^2)^4}\,,\quad f^{\rm SYM}(\lambda)=1+{15 \zeta(3)\over 8} \lambda^{-{3\over 2}} +\cO (\lambda^{-{5\over 2}})\,,
\label{ESYMthooft}
\fe
where $f^{\rm SYM}(\lambda)$ receives contributions from tree-level higher derivative interactions in type IIB string theory.\footnote{A priori one may wonder if there are worldsheet instanton corrections  that depend on the complex moduli $\tau$ of the $T^2$. However there are no nontrivial second homology class in the AdS soliton geometry that can support worldsheet instantons of finite actions. This does not completely rule out worldsheet-instanton-like contributions that are non-perturbative in $\ell_s^2$, equivalently $1\over \sqrt{\lambda}$ (see for example \cite{Tong:2002rq}). Here we have suppressed such potential contributions in \eqref{ESYMthooft}.} 
As the consequence of the relation \eqref{c1rel}, the same function $f(\lambda)$ determines the thermal free energy of the SYM in the 't Hooft limit which has been studied extensively in the literature since \cite{Gubser:1998nz}.  

We end by noting the sharp distinction between the $\tau$ dependence of the toroidal Casimir energy $\cE(\tau)$ in CFTs with matrix large $N$ limits discussed here which have Einstein gravity duals, and those with vector large $N$ limits which have higher-spin gravity duals. In particular, at $d=3$, for the ABJM theory and the $O(N)$ CFT in the leading large $N$ limit, we have 
\ie  
&{\cal E}_{{\rm ABJM}_{N,k}}(\tau) =   {8\sqrt{2}\pi^2\over 81} k^{1/2} N^{3/2}
{A^{3/2}\over {\rm sys} (T^2)^3}\,,
\\
& {\cal E}_{O(N)}(\tau)= N \left(
{2\zeta(3)\over 5\pi} \tau_2^{3/2} + e^{-2\pi\Delta_0 \tau_2} \sqrt{2\pi\Delta_0}\left(
    1+{3\over 16\pi \tau_2 \Delta_0}+\cO(\tau_2^{-2})
\right)
+\cO(e^{-4\pi  \Delta_0\tau_2})
\right) \,.
\fe
The expression for the $O(N)$ CFT is smooth in $\tau \in \mH$ and has an infinite tower of instanton contributions, both of which are features absent in the ABJM case. This signals the different natures of their holographic duals and it would be interesting to investigate this further in the context of ABJ triality \cite{Chang:2017mxc}.

\section{Discussions}
\label{sec:discussion}

In this paper we have studied general properties of $d>2$ CFT on $\cM_d=T^2\times \mR^{d-3,1}$ (or $\cM_d=T^2\times \mR^{d-2}$ in the Euclidean signature) which is one of the simplest spacetime manifold of nontrivial topology and a natural generalization of the thermal background $S^1\times \mR^{d-1}$. We focused on a basic observable on this geometry, namely the Casimir energy (ground state energy) on the spatial manifold $\cM_{d-1}=T^2\times \mR^{d-3}$. The dimensionless Casimir energy density $\cE(\tau)$ is a nontrivial modular invariant function of the complex moduli $\tau$ of the torus $T^2$ and appears to be independent from conventional CFT data.
In the thin torus limit $\tau_2\to \infty$, we derived a simple universal formula \eqref{EFTres} for $\cE(\tau)$ which shows that the toroidal Casimir energy is controlled by two perturbative terms in $\tau_2$ up to non-perturbative corrections. This was accomplished via an effective field theory (EFT) argument by compactifying the $d$ dimensional CFT on the small cycle of the thin torus. The coefficients $c_1,c_2$ of the perturbative terms are proportional to familiar finite temperature observables in the CFT and its circle reduction. The remaining non-perturbative terms in \eqref{EFTres} are accounted for by worldline instantons associated with massive particles in the $d-1$ dimensional EFT going around the remaining cycle in the base manifold. Combining with $PSL(2,
\mZ)$ spectral theory which is a powerful framework to study modular invariant functions, we translated the EFT constraints into a set of stringent conditions on the spectral overlap in the  decomposition of $\cE(\tau)$ with respect to eigenfunctions of the $PSL(2,\mZ)$ invariant Laplacian. This spectral decomposition makes more explicit the class of modular invariant functions that are physically relevant for describing the toroidal Casimir energy $\cE(\tau)$ in CFT. It also explains intricate relations between EFT data that enters into the universal formula for $\cE(\tau)$  \eqref{EFTres}. For example, although $\cE(\tau)$ in general receives worldline instanton contributions in the EFT from all KK charges $Q$, effects at higher total KK charges are completely determined by the $Q=0$ and $Q=\pm 1$ sector as a consequence of the modular invariance. We illustrate these universal properties of $\cE(\tau)$
in concrete CFT examples including the free scalar CFT, the critical $O(N)$ model and holographic CFTs with Einstein gravity duals. Below we discuss a number of open questions and future directions.

\subsubsection*{Sign of Casimir Energy and Universal Bounds}

One intriguing question is the sign of the Casimir energy density and possible universal bounds on its magnitude.
From the universal behavior of the toroidal Casimir energy \eqref{EFTres}, it is clear that $\cE(\tau)$ is positive for large $\tau_2$ and unbounded from above. This translates to a negative vacuum energy density $E_{\rm vac}$ that can be arbitrarily negative by tuning $\tau$.\footnote{Recall  the negative sign in the relation \eqref{Evactodensity} between $E_{\rm vac}$ and $\cE(\tau)$.}
Furthermore, we observe in   all the examples studied here $\cE(\tau)$ is positive everywhere on the upper half-plane. We are thus led to the following conjecture,
\begin{conjecture} 
    For any $d\geq 3$ unitary bosonic CFT (that is not a TQFT), the dimensionless modular-invariant toroidal Casimir energy density $\cE(\tau)$ as defined in \eqref{Evactodensity} is strictly positive.
\end{conjecture}
For $\tau_1=0$, this positivity was proven in  \cite{Belin:2016yll}. In fact in that case a stronger result holds \cite{Belin:2016yll,Levine:2022wos}, stating that along $\tau_1=0$, $\cE(\tau_2) = \epsilon_{\rm vac} \tau_2^{-3/2} (1 + f(\tau_2))$, where $\epsilon_{\rm vac}$ is some positive thermal coefficient, and $f(\tau_2)$ is positive, monotonically increasing and convex in $\tau_2\in (0,\infty)$. 
For 
holographic $d=3$ CFTs, a gravity argument was given in \cite{Hickling:2015tza,Fischetti:2017sut} for the non-positivity of vacuum energy density $E_{\rm vac}\leq 0$ on $\cM_3=\mR\times \Sigma$ with a general closed spatial two-dimensional manifold $\Sigma$ (see also \cite{Galloway:2015ora}). It would be interesting to find a field theoretic proof that applies at general $\tau$ (and potentially for more general closed spatial  manifolds\footnote{Let us collect some field theory evidence for this statement in $d=3$ when the CFT is defined on $\mR \times \Sigma$. When $\Sigma$ is a sphere, we have $E_{\rm  vac}(S^2)=0$ which follows from the operator-state correspondence and the absence of Weyl anomaly in $d=3$ (for the same reason the spherical Casimir energy vanishes for unitary CFT in all odd spacetime dimensions). For a deformed $S^2$, evidence for negative $E_{\rm vac}$ can be found in \cite{Fischetti:2020knf,Cheamsawat:2020awh} for free theories.
For
free scalar CFT when $\Sigma$ is a compact hyperbolic surface (see \cite{Schaden:2005mu} for more general discussions for free massless scalars),  the regulated and rescaled vacuum energy is determined by the Selberg zeta function for the hyperbolic Laplacian $\Delta_\Sigma$ as 
$E_{\rm  vac}(\Sigma) ={1\over 2}\left.\zeta_{\Delta_\Sigma}(s)\right|_{s=-{1\over 2}}$. This zeta function is negative at $s=-{1\over 2}$ 
as a consequence of the Selberg trace formula (see Corollary 3.4 in \cite{kurokawa2002casimir}). Moreover, the  genus two Bolza surface is an extrema (a 
 consequence of its large discrete isometry group \cite{klein2009extremal}) for  $E_{\rm  vac}(\Sigma)$ over the Teichm\"uller space (moduli space of hyperbolic structures) with $E_{\rm  vac}\approx -0.325003$
\cite{Strohmaier_2012}. It would be interesting to show if this is the global maximum (e.g. by generalizing the bootstrap analysis in \cite{Bonifacio:2021aqf,Kravchuk:2021akc} which found that the Bolza surface maximizes the spectral gap).}).

As a consequence of the modular invariance, the special points $\tau=i$ and $\tau=e^{\pi i/3}$ on the standard fundamental domain which are preserved respectively by $\mZ_2$ and $\mZ_3$ subgroups of $PSL(2,\mZ)$ are extrema of $\cE(\tau)$. 
From the CFT examples we have studied, we  notice that  the $\mZ_3$-symmetric point $\tau=e^{2\pi i \over 3}$  is always the global minimum of $\cE(\tau)$ while the $\mZ_2$-symmetric point $\tau=i$ is a saddle point. For free scalar theories, this is a property of the real analytic Eisenstein series $E_{d\over 2}(\tau)$.\footnote{To see this, we note that the Eisenstein series $E_s(\tau)$ is related to the Epstein series $E(\Lambda,s)$ on the two dimensional lattice $\Lambda$ that defines the flat torus via $T^2=\mC/\Lambda$ as below,
\ie  
E(\Lambda,s)\equiv \sideset{}{'}\sum_{v\in \Lambda} \la v,v\ra^{-s}\,,\quad E_s(\tau)= {1 \over2 \zeta(2s)}E(\Lambda,s)\,,
\fe 
where $v=m\tau+n$ and the quadratic form is defined by $\la v,v\ra\equiv  {|m\tau+n|^2\over \tau_2}$ such that the corresponding Gram matrix has determinant one. It is well-known that the minimization of the Epstein series on a rank $d$ lattice is related to finding the densest sphere packing in $\mR^d$ (more precisely lattice type packing). For $d=2$, the minimum of $E(\Lambda,s)$ (and correspondingly the densest sphere packing in $\mR^2$) is achieved  by the triangular lattice which corresponds to $\tau=e^{2\pi i /3}$ 
\cite{rankin1953minimum,cassels1959problem,ennola1964lemma,diananda1964notes,sarnak2006minima}.
}
For the critical $O(N)$ model, these structures are clear from Figure~\ref{fig:ONenergy}.
For holographic CFTs, it follows from Loewner's inequality for the systole on $T^2$ \cite{gromov1983filling},
\ie 
{\rm sys}(T^2)\leq {2\over \sqrt{3}}{\rm vol}(T^2) \,, 
\fe 
which is saturated for flat torus with precisely $\tau=e^{2\pi i\over 3}$.
We thus propose the following conjecture,
\begin{conjecture} 
    For any $d\geq 3$ unitary bosonic CFT (that is not a TQFT), the dimensionless modular-invariant toroidal Casimir energy density $\cE(\tau)$ has a global minimum at $\tau=e^{2\pi i \over 3}$ on the standard $PSL(2,\mZ)$ fundamental domain.
\end{conjecture}

\subsubsection*{Generalizations to Fermionic CFTs}
Thus far we have mostly focused on bosonic CFTs (except for comments in the holographic examples). In the presence of fermionic matter, to define the CFT on a general spacetime manifold $\cM_d$ requires a spin structure. For $\cM_d=T^2\times \mR^{d-2}$, there are four spin structures that correspond to either periodic or anti-periodic boundary conditions for the fermions along each of the two independent cycles on $T^2$. The choice of the spin structure $\rho$ is a part of the data that specifies the fermionic CFT on this geometry (in addition to the metric). The ground state and the corresponding Casimir energy $\cE_{\rho}(\tau)$ depend on $\rho$ in addition to the complex moduli $\tau$. Relatedly, for a fixed spin structure $\rho$, the full $PSL(2,\mZ)$ invariance of the bosonic Casimir energy $\cE(\tau)$ is broken to congruence subgroups preserving $\rho$ for the fermionic counterpart $\cE_{\rho}(\tau)$. Consequently, $\cE_{\rho}(\tau)$ will have different modular properties. There is also a simple generalization of the EFT analysis in Section~\ref{sec:EFT} for fermionic CFT with an analogous universal formula as \eqref{EFTres}. In subsequent work \cite{FermionCFT}, we study the toroidal Casimir energy for fermionic CFTs  systematically with concrete examples including the Gross-Neveu model and Chern-Simons-Matter CFTs.

 \subsubsection*{Excited States and Refinement by Symmetries}

It is also interesting to study excited states in the Hilbert space on $\cM_{d-1}=T^2\times \mR^{d-3}$ and the energy spectrum. In particular, in the similar way that the asymptotic density of high energy states in $d=2$ CFT on $S^1$ is determined by the CFT central charge via the Cardy formula 
\cite{Cardy:1986ie}, 
the asymptotic density of states on $T^{d-1}$ is determined by the Casimir energy using modular invariance on $T^d$ \cite{Belin:2016yll}.\footnote{The case $\cM_{d-1}=T^2\times \mR^{d-3}$ can be thought of a limit of $\cM_{d-1}=T^{d-1}$.} It would be interesting to compute this asymptotic density explicitly in $d\geq 3$ CFTs, such as the 3d $O(N)$ critical model on $\mR\times T^2$ (see \cite{PhysRevB.94.085134}).

In CFTs with global symmetries, it is natural to refine the observables by including  symmetry twists (insertions of topological defects representing the symmetry), which organize the states into representations of the symmetry and also 
give rise to twisted sectors in the Hilbert space. In particular, the Cardy formula in $d=2$ has a symmetry-refined version
\cite{Pal:2020wwd,Lin:2022dhv} that characterizes the asymptotic growth of operator degeneracies in a fixed charge or twist sector. The generalization to higher $d$ CFTs with zero-form symmetries was recently explored in  
\cite{Kang:2022orq}.  
Symmetries are particularly interesting when they carry 't Hooft anomalies, which are often reflected by degeneracies in the twisted Hilbert space \cite{Delmastro:2021xox}. For $d=3$ CFT on $\cM_3=\mR \times T^2$, the spatial manifold supports both nontrivial zero-form and one-form symmetry defects and thus provides an ideal playground to investigate consequences of anomalies for generalized symmetries (and gravity)  in CFT.

 \subsubsection*{State-operator Correspondence on $T^2$ and Line Defects}

In contrast to the familiar  correspondence between states on $S^{2}$ and local operators in a 3d CFT, the state-operator correspondence on a spatial $T^2$ is much more mysterious. In particular, as explained in \cite{Belin:2018jtf}, a basic difficulty comes from the obstruction in realizing the ground state on $T^2$ via a Euclidean path integral over a compact three-manifold that fills in the $T^2$. It is natural to expect line defects to play an important role in the construction of states on $T^2$. After all, this is how states on $T^2$ in a 3d TQFT are constructed. They come from anyons threading the non-contractible cycle of the solid torus that fills in the $T^2$ \cite{Witten:1988hf}. For general 3d CFTs, while it is not clear what states on $T^2$ can be constructed by threading a non-topological line defect in the solid torus, one can turn the question around and  deduce 
constraints on the line defects in the CFT. This is currently under investigation.

\section*{Acknowledgments} 

We thank Nathan Benjamin, Scott Collier, Himanshu Khanchandani, Zohar Komargodski, Juan Maldacena, and Edgar Shaghoulian for interesting discussions. We are also grateful to Nathan Benjamin and Edgar Shaghoulian for helpful comments on a draft. The work of YW was
supported in part by NSF grant PHY-2210420.

\appendix

\section{Regularization of Lattice Sum on Torus}
\label{app:Gfunc}

In the main text, we have encountered the following sum over lattice momentum vectors on a torus of complex moduli $\tau$,
\ie  
\sum_{m,n\in \mZ}{1\over (|m\tau+n|^2+\tau_2^2\Delta(\tau)^2)^{s}}\,.
\label{tbr}
\fe 
The sum with $s=-\frac{1}{2}$ appears in the expressions for the Casimir energy on $T^2$, and the sum with $s=\frac{1}{2}$ appears in the gap equation (e.g. for the $O(N)$ model at large $N$). This sum is convergent for $s>1$, and will require regularization otherwise. 
In this appendix we discuss regularization of such sums using a generalization of the real analytic Eisenstein series.


In \cite{PhysRevB.94.085134}, a regularization of \eqref{tbr} is provided by analytic continuation in both $s$ and $d$ (the rank of the lattice) via the integral representation in the last equality below,
\ie
& g_s^{(d)}(\Delta, \tau,a_1,a_2) = \sum_{n, m \in \mathbb{Z}^{d / 2}} \frac{1}{\left(\left|m+a_2+\left(n+a_1\right) \tau\right|^2+\gamma^2\right)^s} \\
&= \frac{\pi^s}{\Gamma(s)} \int_0^{\infty} d \lambda \lambda^{s-1} \exp \left[-\pi \lambda \gamma^2-\frac{d \pi \lambda}{2}\left(\left(a_1 \tau_2\right)^2+\left(a_2+a_1 \tau_1\right)^2\right)\right] \Theta\left(\lambda, \boldsymbol{\Omega}(\tau), \mathbf{v}_1\right)^{d / 2} \\
&= \frac{\pi^s}{\Gamma(s)}\left\{\int_1^{\infty} d \lambda \lambda^{s-1} \exp \left[-\frac{\lambda \tau_2^2 L^2 \Delta^2}{4 \pi}-\frac{d \pi \lambda}{2}\left(\left(a_1 \tau_2\right)^2+\left(a_2+a_1 \tau_1\right)^2\right)\right] \Theta\left(\lambda, \boldsymbol{\Omega}(\tau), \mathbf{v}_1\right)^{d / 2}\right. \\
&+\tau_2^{-d / 2} \int_1^{\infty} d \lambda \lambda^{d / 2-s-1}\left[\exp \left(-\frac{\tau_2^2 L^2 \Delta^2}{4 \pi \lambda}\right) \Theta\left(\lambda, \boldsymbol{\Omega}(\tau)^{-1}, \mathbf{v}_2\right)^{d / 2}-1+\frac{\tau_2^2 L^2 \Delta^2}{4 \pi \lambda}\right] \\
&\left.+\frac{\tau_2^{-d / 2}}{s-d / 2}-\frac{L^2 \Delta^2}{4 \pi} \frac{\tau_2^{2-d / 2}}{1+s-d / 2}\right\}\,,
\label{WSform}
\fe
where
\ie
& \gamma = \frac{\tau_2 L \Delta}{2\pi}, \quad \boldsymbol{\Omega}(\tau)=\left(\begin{array}{cc}
|\tau|^{2} & -\tau_{1} \\
-\tau_{1} & 1
\end{array}\right), \quad \boldsymbol{\Omega}(\tau)^{-1}=\frac{1}{\tau_{2}^{2}}\left(\begin{array}{cc}
1 & \tau_{1} \\
\tau_{1} & |\tau|^{2}
\end{array}\right), \\
&\mathbf{v}_{1}=t\left(\begin{array}{c}
\tau_{1}\left(-a_{2}+a_{1} \tau_{1}\right)+a_{1} \tau_{2}^{2} \\
a_{2}-a_{1} \tau_{1}
\end{array}\right), \quad \mathbf{v}_{2}=-i\left(\begin{array}{l}
a_{1} \\
a_{2}
\end{array}\right), \\
& \Theta(t, \boldsymbol{\Omega}, \mathbf{u}) \equiv \sum_{\mathbf{n} \in \mathbb{Z}^{2}} \exp \left(-\pi t \mathbf{n}^{\top} \cdot \boldsymbol{\Omega} \cdot \mathbf{n}-2 \pi \mathbf{n}^{T} \cdot \mathbf{u}\right) .
\label{gs_notation}
\fe
Here $a_1,a_2\in [0,1)$ keep track of possible twisted boundary conditions for a complex scalar along the two independent cycles of $T^2$.

The regulated expression above from \cite{PhysRevB.94.085134} is finite at $d=2$ and $s=\pm 1/2$ which are the cases relevant for the Casimir energy and gap equation in the   $O(N)$ scalar CFT. However we find the integral formula cumbersome to deal with both analytically and numerically.

Here we derive alternative but equivalent representations of the 
regulated lattice sum which will be more efficient for our analysis and 
 will also make more explicit the modular properties. We start by defining the following 
generalized Eisenstein series,
\ie 
G_{s}(\tau,\Delta,a_1,a_2)=\frac{1}{2} \pi^{-s} \Gamma(s) \sum_{n,m\in \mathbb{Z}} \left( \frac{\tau_2}{|(m+a_2) - (n+a_1)\tau|^2 + \Delta^2 \tau_2^2} \right)^s \,,
\label{GSapp}
\fe 
where the extra $\tau_2^s$ factor compared to \eqref{WSform} is introduced so that $G_s$ transforms nicely under $PSL(2,\mZ)$.\footnote{\label{fn:conv}Note that in going between \eqref{WSform} from \cite{PhysRevB.94.085134} and our \eqref{GSapp}, there is a notation change $L^{\rm there}=L_1^{\rm here}$.}
In this paper, we only make use of $G_s(\tau,\Delta,a_1,a_2)$ with $a_1=a_2=0$ \eqref{Gsdef}
which is a modular invariant function as long as $\tau_2 \Delta^2(\tau)$ is modular invariant. For general $a_i$, $G_s(\tau,\Delta,a_1,a_2)$ transforms as a non-holomorphic Jacobi form. In the following we keep $a_i$ general since the manipulations 
we present work uniformly regardless of the values of $a_i$
and the general case will be useful when incorporating symmetry twists in subsequent work.



We will offer two different regulated expressions for the 
generalized Eisenstein series $G_s$ defined in \eqref{GSapp}, which are complementary and will both be useful for different situations.  One is given by a sum of Poincar\'e series,
\ie
\label{g_function}
  G_s(\tau, \Delta, a_1, a_2)  
= &\sideset{}{'}\sum_{m,n \in \mathbb{Z}} (\frac{\tau_2 \Delta}{|n+m\tau|})^{1-s} K_{1-s}(2\pi \Delta |n+m\tau|) e^{2\pi i(a_1 n+a_2 m)} \\
&+ \frac{\Gamma(s-1)}{2} (\pi \tau_2 \Delta^2)^{1-s} \,,
\fe
and the other in terms of a Fourier series in $\tau_1$,\footnote{Note that $\Delta$ still has dependence on $\tau_1$ in general.}
\ie
\label{bessel_sum}
   G_{s}(\tau, \Delta, a_1, a_2) 
=& \sum_{n \in \mathbb{Z} + a_1} \sum_{m \neq 0} e^{2\pi i m(n\tau_1 - a_2)} |m|^{s-{1\over 2}} \tau_2^{1\over 2}\left( \Delta^2 + n^2 \right)^{\frac{1-2s}{4}} K_{\frac{1}{2} - s} (2\pi \sqrt{\Delta^2 + n^2} |m| \tau_2) \\
&+ 2 \tau_2^{1-s} \sum_{n \in \mathbb{Z}^+} \left(\frac{\Delta}{n}\right)^{1-s} K_{1-s}(2\pi \Delta n) \cos(2\pi n a_1) + \frac{\Gamma(s-1) }{2} (\pi \tau_2 \Delta^2)^{1-s} \,.
\fe

We first derive \eqref{g_function} from the expression in the second line of \eqref{WSform} (see footnote~\ref{fn:conv} for the notation change) by performing Poisson resummations in both $n$ and $m$ for the modes  of the Riemann theta function $\Theta$ labeled by ${\bf n}=(n,m)$, and then further splitting the new sum into  the nonzero modes (${\bf n}\neq (0,0)$) and 
 the zero mode (${\bf n}=(0,0)$),
\ie
& \, G_s(\tau, \Delta, a_1, a_2) = \, \frac{\tau_2^{s-1}}{2} \int_{0}^{\infty} dt t^{-s} \exp \left( -\frac{\pi \tau_2^2 \Delta^2}{t} \right) \left( \Theta\left(t, \boldsymbol{\Omega}(\tau)^{-1}, \mathbf{v}_{2}\right) -1 \right) 
\\
+& \frac{\tau_2^{s-1}}{2} \int_0^{\infty} dt t^{s-2} e^{-\pi t \tau_2^2 \Delta^2}  \,.
\fe
The integral of the nonzero modes above produce the sum of Bessel functions in \eqref{g_function}. The integral of the zero mode is regulated by analytic continuation in $s$, yielding the remaining term in \eqref{g_function}.

The other regulated formula \eqref{bessel_sum} for the generalized Eisenstein series is derived 
by Poisson resummation in a different way. We split the Riemann theta function $\Theta$ into modes ${\bf n}=(n,m)$ with $m=0$ or $m\neq 0$, and perform Poisson resummation with respect to $n$. The zero mode is regulated by analytic continuation in $s$ as above. The resulting expression in \eqref{bessel_sum} then follows.

\bibliographystyle{JHEP}
\bibliography{torus}

\end{document}